\documentclass[a4paper,fleqn,usenatbib]{mnras}
\usepackage{cleveref}
\usepackage{graphicx}
\usepackage{booktabs}
\usepackage{multirow}
\usepackage{color}
\usepackage[T1]{fontenc}
\usepackage{hyperref}
\usepackage{ae,aecompl}

\newcommand{\tabitem}{~~\llap{\textbullet}~~}
\definecolor{darkgreen}{rgb}{0.0, 0.2, 0.13}

\begin{centering}\title[Modeling Galaxy Formation and Evolution by using Machine Learning]{Machine Learning and Cosmological Simulations I: Semi-Analytical Models}
\author[Harshil M. Kamdar, Matthew J. Turk, Robert J. Brunner]{Harshil M. Kamdar$^{1,2}$\thanks{E-mail:
hkamdar2@illinois.edu}, Matthew J. Turk$^{2,4}$ and Robert J. Brunner$^{2,3,4,5}$ \\
$^{1}$Department of Physics, University of Illinois, Urbana, IL 61801 USA \\
$^{2}$Department of Astronomy, University of Illinois, Urbana, IL 61801 USA \\ 
$^{3}$Department of Statistics, University of Illinois, Champaign, IL 61820 USA \\
$^{4}$National Center for Supercomputing Applications, Urbana, IL 61801 USA \\
$^{5}$Beckman Institute For Advanced Science and Technology, University of Illinois, Urbana, IL, 61801 USA}
\end{centering}
\begin{document}

\date{Accepted 2015 October 1. Received 2015 September 30; in original form 2015 July 2}

\pagerange{\pageref{firstpage}--\pageref{lastpage}} \pubyear{2015}

\maketitle

\label{firstpage}

\begin{abstract}
We present a new exploratory framework to model galaxy formation and evolution in a hierarchical universe by using machine learning (ML). Our motivations are two-fold: (1) presenting a new, promising technique to study galaxy formation, and (2) quantitatively analyzing the extent of the influence of dark matter halo properties on galaxies in the backdrop of semi-analytical models (SAMs). We use the influential Millennium Simulation and the corresponding Munich SAM to train and test various sophisticated machine learning algorithms (k-Nearest Neighbors, decision trees, random forests and extremely randomized trees). By using only essential dark matter halo physical properties for haloes of $M>10^{12} M_{\odot}$ and a partial merger tree, our model predicts the hot gas mass, cold gas mass, bulge mass, total stellar mass, black hole mass and cooling radius at z = 0 for each central galaxy in a dark matter halo for the Millennium run. Our results provide a unique and powerful phenomenological framework to explore the galaxy-halo connection that is built upon SAMs and demonstrably place ML as a promising and a computationally efficient tool to study small-scale structure formation.
\end{abstract}


\section{Introduction} \label{mgf:intro}
\par
In recent years, with the introduction of surveys such as SDSS\footnote{www.sdss.org}, DES\footnote{www.darkenergysurvey.org}, and LSST\footnote{www.lsst.org}, the amount of data available to astronomers has exploded. These massive data sets have enabled astronomers to form and test sophisticated models that explain cosmic structure formation in the universe. Cosmological simulations are a rich subset of these models and have consequently, also been on the rise; these simulations provide a concrete link between theory and observation. It has been argued that the $\Lambda$CDM model ~\citep{peebles1982large, blumenthal1984formation, davis1985evolution} is as widely accepted as it is today largely due to the emergence of these high-resolution numerical simulations \citep{springel2005cosmological}. However, modeling galaxy formation accurately by using numerical simulations remains an important problem in modern astrophysics, both scientifically and computationally. 
\par
	The evolution of collisionless dark matter particles at large scales has been studied exhaustively at unprecedentedly high resolutions, given the meteoric rise in computational power and the relative simplicity of these simulations \citep{springel2005cosmological, springel2005simulations, klypin2011dark, angulo2012scaling, skillman2014dark}. The formation of structure on the scale of galaxies, however, has been incredibly difficult to model \citep{somerville2014physical}; the difficulty arises primarily because baryonic physics at this scale is governed by a wide range of dissipative and/or nonlinear processes, some of which are poorly understood \citep{kang2005semianalytical, baugh2006primer, somerville2014physical}. 
    
\par Broadly speaking, there are two prevalent techniques used to understand galaxy formation and evolution: semi-analytical modeling (SAM) and simulations that include both  hydrodynamics and gravity. The former is a post de facto technique that combines dark matter only simulations with approximate physical processes at the scale of a galaxy \citep{baugh2006primer}. The SAM used in this work is detailed in \citet{croton2006many, de2006formation, de2007hierarchical} (hereafter DLB07), and \citet{guo2011dwarf} (hereafter G11). For a general, exhaustive review of the motivation of SAMs and a comparison of different SAMs, the reader is referred to \citet{baugh2006primer, somerville2014physical} and \citet{knebe2015nifty}. N-body + hydrodynamical simulations (NBHS) evolve baryonic components using fluid dynamics alongside regular dark matter evolution. The biggest advantage of NBHS over SAMs is the self-consistent way in which gaseous interactions are treated in the former. However, NBHS are incredibly computationally expensive to run and also require some approximations at the subgrid level similar to those applied in SAMs. Promising new NBHS are outlined in \citet{vogelsberger2014introducing} and \citet{schaye2015eagle}. For an extensive comparison of SAMs and NBHS, the reader is referred to \citet{benson2001comparison, yoshida2002gas, monaco2014semi} and \citet{somerville2014physical}.

\par
Dark matter plays an integral role in galaxy formation; broadly speaking, dark matter haloes are `cradles' of galaxy formation \citep{baugh2006primer}. It is well-established that gas cools hierarchically in the centers of dark matter haloes through mergers; the evolution of galaxies, however, is dictated by a wide variety of baryonic processes that are discussed later in this paper. While baryonic physics plays a crucial role in the outcome of gaseous interactions, the story always starts with gravitational collapse. However, no simple mapping has been found between the internal dark matter halo properties and the final galaxy properties because of the sheer complexity of the baryonic interactions. For instance, in \citet{contreras2015galaxy}, a systematic study of the relationship between the host halo mass and internal galaxy properties is performed. They conclude that no simple mapping was found between the cold gas mass or the star formation rate and the host halo mass. The lack of a relatively simple mapping between internal halo properties and the galaxy properties motivates many of the approximations that SAMs and NBHS make. 

\par Moreover, the computational costs associated with both standard galaxy formation models are incredibly high. The Illustris simulation (an NBHS) used a total of around 19 million CPU hours to run.\footnote{http://www.illustris-project.org/about/} SAMs, while significantly faster than NBHS, still require an appreciable amount of computational power. For instance, consider the open source GALACTICUS SAM put forth in \citet{benson2012galacticus}; in GALACTICUS, a halo of mass $10^{12} M_{\odot}$ is evolved (with baryonic physics) in around 2 seconds and a halo of mass $10^{15} M_{\odot}$ is evolved in around 1.25 hours. Thus, a very rough order of magnitude estimate can be made for the approximate runtime for GALACTICUS. For about 500,000 dark matter haloes, and an average evolution time of approximately 2 minutes (corresponding to about $10^{13} M_{\odot}$), the time taken for GALACTICUS to build merger trees to $z=0$ is $O(15000)$ CPU hours. The lack of a simple mapping between dark matter haloes and the properties of galaxies, the computational costs associated with the popular galaxy formation models and the highly nonlinear nature of the problem make galaxy formation incredibly hard to model, leaving room for new exploration.

\par 
While SAMs, not limited to DLB07 and G11, have been incredibly successful in reproducing a lot of observations \citep{white1991galaxy, kauffmann1993formation, cole1994recipe, somerville1999semi, cole2000hierarchical, kang2005semianalytical, bower2006breaking, monaco2007morgana, de2007hierarchical, lagos2008effects, somerville2008semi, weinmann2010cluster, de2011comparison} and produce similar results to NBHS \citep{benson2001comparison, somerville2014physical}, there still exist a few deficiencies in the general methodology of SAMs. Most importantly, the degeneracy inherent to most SAMs is concerning (see, for e.g., \citet{henriques2009monte, bower2010parameter, neistein2010degeneracy}). SAMs (including DLB07 and G11) use simple yet powerful, physically motivated analytical relationships for most processes that play a role in galaxy formation; these processes have several free parameters that are `tuned' to match up with observations. 

\par An alternative approach to model galaxy formation, that is physically much more transparent, was employed in \citet{neistein2010degeneracy} (hereafter referred to as NW). NW put forth a simple model that includes treatment of feedback, star formation, cooling, smooth accretion, gas stripping in satellite galaxies, and merger-induced starbursts with one key difference compared to conventional SAMs. In the NW model, the efficiency of each physical process is assumed to depend only on the host halo mass and the redshift, making it a much simpler model than G11, DLB07, and other SAMs. NW produces a very similar population of galaxies with similar physical properties to that of DLB07's (G11's predecessor). The success of NW raises an interesting question: could we go even further and try to learn more about the halo-galaxy connection using solely the halo environment and merger history, and would we be able to reproduce results found in conventional SAMs? However, attempting to do so is a non-trivial task for several reasons. First, the inputs for the exploratory model aren't exactly clear. Second, the mappings between dark matter halo properties and galaxy properties are incredibly complex, as discussed earlier. NW, G11, and all other SAMs use simplified analytical relationships to capture complex baryonic processes; these relationships have a partial, non-trivial dependence on internal halo properties but it is not clear how these analytical relationships can be used to build a dark matter-only model to probe galaxy formation and evolution. Given their non-parametric nature and their ability to successfully model complex phenomena, machine learning algorithms provide an interesting framework to explore this problem. 

\par
A variety of statistical techniques, falling under the broad subfield of machine learning (ML), are gaining traction in the physical sciences. The main goal of ML is to build highly efficient, non-parametric algorithms that attempt to learn complex relationships in and make predictions on large, high-dimensional data sets. Applications of ML to model highly complex physical models include pattern recognition in meteorological models \citep{liu2011review}, particle identification \citep{roe2005boosted}, inferring stellar parameters from spectra \citep{fiorentin2007estimation}, photometric redshift estimation \citep{kind2013tpz} and source classification in photometric surveys \citep{kim2015hybrid}. Machine learning techniques have been shown to be highly effective at picking up complex relationships in high-dimensional data \citep{witten2005data, johnson2011learning, graff2014skynet}. As discussed later, we find that ensemble techniques that use a combination of decision trees perform the best in the context of galaxy formation. The relative simplicity of some ML techniques, their high computational efficiency, and powerful predictive capabilities for complex models make the problem of galaxy formation well-suited problem for machine learning.

\par 
In this paper, we present a first exploration into using supervised ML techniques to model galaxy formation. For our analyses, we use the high-resolution Millennium simulation \citep{springel2005cosmological, springel2005simulations} performed by the Virgo Consortium. The Millennium simulation is an extremely influential dark matter simulation that has motivated more than 700 papers in the study of large scale structure and galaxy evolution. The nature of the Millennium simulation makes it ideal for galaxy-scale studies. The internal halo properties and a partial merger history for each dark matter halo at $z=0$ in the Millennium simulation are extracted to be used as input features for our algorithms. We use the well-established Munich SAM, G11, for Millennium \citep{croton2006many, de2006formation, de2007hierarchical, guo2011dwarf} for our training and testing. We use various ML algorithms to predict the cold gas mass, hot gas mass, stellar mass in the bulge, total stellar mass and the central black hole mass for the central galaxy of each dark matter halo in the Millennium simulation. These components of the mass at $z=0$ provide an extensive probe into how effective ML algorithms are at learning baryonic processes prescribed in SAMs by using only dark matter inputs. 

\par
It should be emphasized here that absolutely no baryonic processes are explicitly included in our analyses. Only the relevant internal dark matter halo properties: number of dark matter particles, spin, $M_{crit200}$, velocity dispersion ($\sigma_v$), maximum circular velocity ($v_{max}$), and a partial merger history of the dark matter halo in which the galaxy resides are used as the inputs for the algorithms. The results of this study will shed valuable light on the halo-galaxy connection. We can quantitatively, admittedly phenomenologically and not physically, determine the extent of the impact that dark matter has on the structure formation at smaller scales in the universe. We can also evaluate how well ML can learn the physical prescriptions that are used in SAMs. To reiterate, our model uses only the `skeleton' of most galaxy formation models: the merger tree of each dark matter halo. 

\par
The paper is organized as follows. In Section \ref{mgf:db}, we discuss the data extracted from the Millennium simulation and the basics of machine learning. More specifically, we outline the basics of the Millennium simulation and G11 and our reasons for using the Millennium simulation. We also discuss the basic principles of ML and outline and discuss the best algorithm that was found empirically. In Section \ref{results}, we outline our results. In this section, the results for the different components of the central galaxy mass are presented. We also discuss the drawbacks of our model and provide possible explanations for some of the discrepancies in our results and present an alternative ML approach to correct for these discrepancies. In Section \ref{discussion}, we discuss what our results imply about the halo-galaxy connection and SAMs. Finally, in Section \ref{conclusion}, we conclude the paper with a summary of our findings and potential avenues for future research. 

\section[]{Data \& Background} \label{mgf:db}
In this section, we discuss the data set obtained by using the Millennium simulation and the ML algorithms that were used for the analyses. First, we discuss general details of the Millennium simulation and the cosmogony that was employed in carrying out the simulation. Next, we discuss the SAM used in Millennium (G11) and give a brief overview of how key physical processes are handled in the model. We also discuss our reasons for choosing the Millennium simulation in place of a higher resolution simulation with a more accurate cosmogony; in particular, Millennium's $\sigma_8$ value is noticeably off from the most recent Planck results \citep{planck2015planck}. We also discuss the extraction of the data set and outline the challenges that were faced in constructing it. Finally, we briefly review how ML works, and outline the primary algorithms were used in our analyses. 
\subsection{Millennium Simulation} \label{mgf:db_mill}
\par The data for this project were extracted from the publicly available Millennium simulation \citep{springel2005cosmological, springel2005simulations}. The Millennium simulation was ran with a custom version of GADGET-2, using the Tree-PM method \citep{xu1994new} to handle gravitational interactions. The Millennium halo catalogs were generated by using a friends-of-friends (FoF) algorithm with a linking length of 0.2 times the mean dark matter particle separation. The Millennium simulation is run with $2160^3$ dark matter particles in a $500 h^{-1} \textrm{Mpc}$ box from $z=127$ to $z=0$. The mass of each dark matter particle is $8.6 \times 10^8 \textrm{M}_{\odot}\textrm{h}^{-1}$ and the smallest subhalo has at least 20 particles. The cosmological model employed in the Millennium simulation has $\Omega_m = 0.25$, $\Omega_b = 0.045$, $\Omega_{\Lambda} = 0.75$, $h = 0.73$, $n_s = 1$ and $\sigma_8 = 0.9$, where the Hubble constant is parametrized as $H_0 = 100 $ km s$^{-1}$ Mpc$^{-1}$. 

\par The raw simulation was sampled in the form of a snapshot 64 times, with FoF group catalogs and their substructures, identified by using SUBFIND, which is discussed in \citet{springel2001populating}. With SUBFIND, each FoF group is decomposed into a set of subhaloes by identifying locally overdense, gravitationally bound regions. The merger tree organization of the dark matter haloes is shown in Figure 11 of \citet{springel2005simulations}. 

\subsection{G11} \label{mdf:db_sam}

\par The SAM used to populate the dark matter haloes in Millennium with galaxies is described extensively in \citet{springel2005simulations, croton2006many, de2006formation, de2007hierarchical, guo2011dwarf}. We only provide a brief overview here. G11 includes ingredients and methodologies originally introduced by \citet{white1991galaxy} and later refined by \citet{springel2001populating, de2004chemical, croton2006many}. G11, like most SAMs, has simple, yet physically powerful prescriptions for gas cooling, star formation, supernova feedback, galaxy mergers, and chemical enrichment that are tuned by using observational data. G11 uses the \citet{chabrier2003galactic} IMF. Additionally, G11 takes into account the growth and activity of central black holes and their effect on suppressing the cooling and star formation in massive haloes. Morphological transformation of galaxies and processes of metal enrichment are also modeled. For a more thorough description of the physical prescriptions used in G11, the reader is referred to the set of papers referenced above. \citet{liu2010stellar, de2011comparison, cucciati2012comparison} show where DLB07 (G11's predecessor) agrees with observational data and also show  some weaknesses of DLB07. Furthermore, \citet{knebe2015nifty} discuss how L-GALXIES (the code behind G11; \citet{henriques2013simulations}) and DLB07 perform against other recent SAMs (e.g. Galacticus, GalICS, etc.) In this section, we provide an overview of how G11 handles the cold gas mass (\ref{mdf: db_sam_cold}), central black hole mass (\ref{mdf: db_sam_bh}), total stellar mass (\ref{mdf: db_sam_stellar}), bulge mass (\ref{mdf: db_sam_bulge}) and the hot gas mass (\ref{mdf: db_sam_hot}).

\subsubsection{Cold Gas Mass} \label{mdf: db_sam_cold}
\par
We outlined in the introduction how the dark matter merger history forms the `skeleton' of SAMs. However, the structure of dark matter haloes and their internal properties are also important in determining the rate at which gas cools and the dynamics of the galaxies in the halo \citep{baugh2006primer}. The cooling of gas in G11 is computed using the growth of the cooling radius $r_{cool}$ as defined in \citet{croton2006many, guo2011dwarf}, which describes the maximum radius at which the hot gas density is still high enough for the cooling to occur within the halo dynamical time $t_h$, following the simple model presented in \citet{springel2001populating}. In G11, it is assumed that infalling gas is shock-heated to the virial temperature ($T_{vir}$) of the dark matter halo at an accretion shock. The cooling time ($t_{cool}$) at each radius is given by:

\begin{equation}
t_{cool} = \frac{3}{2} \frac{\mu m_p k T}{\rho_{hot}(r) \Lambda (T_{hot},Z_{hot})}
\end{equation}

Here, $\mu m_p$ denotes the mean particle mass, $k$ is the Boltzmann constant, $\rho_{hot}(r)$ is the hot gas density, $\Lambda(T_{hot}, Z_{hot})$ is the cooling function, that depends on temperature and metallicity \citep{sutherland1993cooling}. $T_{hot}$ is given by: 
\begin{equation}
T_{hot} = 35.9 \left(\frac{V_{vir}}{km s^{-1}}\right)^2 
\end{equation}
The cooling radius as the point where the local cooling time is equal to $t_h$, where:
\begin{equation}
t_h = \frac{R_{vir}}{V_{vir}} = 0.1 H(z)^{-1}
\end{equation}
Therefore, the cooling radius can be written as: 
\begin{equation}
r_{cool} = \left[\frac{t_{dyn} m_{hot} \Lambda (T_{hot},Z_{hot})}{6 \pi \mu m_h k T_{vir} R_{vir}}\right]^{\frac{1}{2}}
\label{rcool}
\end{equation}
The cooling rate can then be written through a simple continuity equation, assuming an isothermal distribution:
\begin{equation}
\dot M_{cool} = \frac{1}{2} \frac{M_{hot} r_{cool} V_{vir}}{R_{vir}^2}
\label{cool}
\end{equation}
\par
A major modification in the cooling rate in Equation \ref{cool} comes about through `radio mode' AGN feedback. AGN feedback becomes especially important in haloes with larger masses. G11 follows the prescription laid out in \citet{croton2006many} for the suppression of this cooling rate; the modified rate is given by:
\begin{equation}
\dot M_{cool}^{'} = \dot M_{cool} - 2 \frac{\dot E_{radio}}{V_{vir}^2}
\label{mod_cool}
\end{equation}
where:
\begin{equation}
\dot E_{radio} = 0.1 \dot M_{BH} c^2
\end{equation}
\begin{equation}
\dot M_{BH} = \kappa \left(\frac{f_{hot}}{0.1}\right) \left(\frac{V_{vir}}{200 kms^{-1}}\right)^3 \left(\frac{M_{BH}}{10^8 M_{\odot} h^{-1}}\right) M_{\odot} yr^{-1}
\end{equation}
Here, $f_{hot}$ for a main subhalo is given by the ratio of the hot gas mass to the subhalo mass ($\frac{M_{hot}}{M_{DM}}$) and $\kappa$ sets the efficiency of the accretion of the hot gas. A more detailed explanation can be found in \citet{guo2011dwarf}. 

\par The rate of gas cooling is an integral part of galaxy formation because it determines the rate at which stars form in a galaxy \citep{baugh2006primer}. As we can see, G11 has a complex recipe to incorporate gas cooling and, while we can see some halo dependence in Equation \ref{cool} and \ref{mod_cool}, this is not a trivial mapping. In the NW model discussed in the introduction, because of the complexity inherent to gas cooling, the coefficient for the cooling rate was empirically determined by running DLB07 on the milli-Millennium simulation since no fitting function in terms of the host halo mass and redshift was found. 
 
\subsubsection{Central Black Hole Mass} \label{mdf: db_sam_bh}
\par G11, like \citet{croton2006many}, splits AGN activity into `quasar' mode and `radio' mode. The formation and evolution of the black hole is dominated by the `quasar' mode feedback, where the central black hole grows through major and/or gas-rich mergers. During a merger, the central black hole of the larger progenitor absorbs the minor progenitor's black hole, and cold gas is accreted onto the central black hole. The evolution of the black hole mass through `quasar' mode feedback is given by:
\begin{equation}
\delta M_{BH} = M_{BH,min} + f\left(\frac{M_{minor}}{M_{major}}\right) \left(\frac{M_{cold}}{1 + \frac{280 km s^{-1}}{V_{vir}}}\right) 
\label{quasar}
\end{equation}
Here, $M_{BH,min}$ is the mass of the black hole in the minor progenitor, $f$ is a free parameter that is set to 0.03 to reproduce the observed BH mass-bulge mass relation, $M_{minor}$ and $M_{major}$ are the total baryonic masses of the minor and major progenitors and $M_{cold}$ is the total cold gas mass.
 
\subsubsection{Stellar Mass} \label{mdf: db_sam_stellar}
\par G11 follows the \citet{kauffmann1996disc} recipe in assuming that star formation is proportional to the mass in cold gas above a certain threshold. A threshold surface density: \begin{equation} \Sigma _{crit} = 12 \times \left(\frac{V_{vir}}{200kms^{-1}}\right) \left(\frac{R}{kpc}\right)^{-1} M_{\odot} pc^{-2} \end{equation} is set for cold gas below which stars do not form and above which stars do form \citep{kennicutt1998global}. Furthermore, it is assumed that this cold gas mass is distributed uniformly over the disk, giving us the following for the cricitcal mass:
\begin{equation}
M_{crit} = 11.5 \times 10^9 \left(\frac{V_{max}}{200kms^{-1}}\right)\left(\frac{r_{disk}}{10 kpc}\right) M_{\odot}
\end{equation} 
\par Therefore, when the mass of the cold gas in the galaxy is greater than $M_{crit}$, the stars form at the following rate per unit time:
\begin{equation}
\dot M_{\star} = \alpha _ {sf} \frac{M_{cold} - M_{crit}}{t_{dyn, disk}}
\end{equation} 
where the disk dynamical time is given by $t_{dyn,disk}=\frac{r_{disc}}{V_{vir}}$    and $\alpha _ {sf}$ is the star formation efficiency, which is manually set between 5 and 15 percent. In the NW model, a modified star formation rate is used:
\begin{equation}
\dot M_{\star} = f_{s} (M_{cold} - M_{crit}) 
\end{equation}
where $f_s$ is the star formation efficiency and has the units of Gyr$^{-1}$. The analytic fitting function that NW found for $f_s$ was given as: 
\begin{equation}
f_s = M_h ^ {1.04} t^{-0.82} 10^{-6.5-0.0394(log(M_h))^2}
\end{equation}
and $M_{crit}$ was parameterized in terms of the host halo mass as: 
\begin{equation}
M_{crit} = f_s ^ {-1} 10^{-8.61} M_h ^ {0.68} t^{-0.515}
\end{equation}
$M_h$ has the units of $M_{\odot}$h$^{-1}$ and t is in $Gyr$. The results presented in Neistein \& Weinmann for stellar mass show good agreement with DLB07 results. The above parametrization of the star formation efficiency and $M_{crit}$ in terms of solely the host halo mass and time and its success offer motivation for our study.

\subsubsection{Bulge Mass} \label{mdf: db_sam_bulge}

\par
In G11, bulge growth is modeled in 3 ways: minor mergers, major mergers and disk instabilities. In the case of minor mergers (i.e. a satellite merging with a  central galaxy), the total stellar mass of the satellite galaxy is added to the bulge of the central galaxy and the disk of the larger progenitor remains intact. The cold gas of the satellite galaxy is added to the disk of the central galaxy and a fraction of the combined cold gas from both galaxies is turned into stars as a result of the merger. In the case of major mergers, the disks of both merging galaxies are destroyed to  form a spheroid to which the combined stellar mass of the two progenitors is assigned. 

\par G11 uses energy conservation and the virial theorem to calculate the change in size:
\begin{equation}
C\frac{GM_{new,bulge}^2}{R_{new,bulge}} = C\frac{GM_{1}^2}{R_{1}} + C\frac{GM_{2}^2}{R_{2}} + \alpha\frac{GM_{1}M_{2}}{R_{1} + R_2}
\end{equation}
Here, $C$ is a parameter relating the binding energy of a galaxy to its mass and radius, and $\alpha$ is a parameter signifying the effective interaction energy in the stellar components.
Furthermore, in G11 a prescription for bulge formation through disk instability is also included. Following the framework set up in \citet{mo1998formation}, it is assumed that a stellar disk becomes unstable when:
\begin{equation}
\frac{V_c}{(\frac{Gm_d}{r_d})^{\frac{1}{2}}} \leq 1
\end{equation}
where $V_c$ is approximated as $V_{vir}$. At each time step, this inequality is checked and if it is not satisfied, some stellar mass is transferred from the disk to the bulge until stability is restored. 

\subsubsection{Hot Gas Mass} \label{mdf: db_sam_hot}
The total hot gas mass is a result of various physical processes. If we ignore supernovae feedback and gas stripping, we obtain the following amount of hot gas for a halo at each snapshot available for cooling:
\begin{equation}
M_{hot} = f_b M_{vir} - \sum_i M_{cold}^{(i)}
\label{hot_analytic}
\end{equation}
Here, the cold gas mass is summed over in all the galaxies in the FOF group and $f_b$ is the universal baryon fraction, given by: 0.017. 
\par Supernova feedback is the main source of reheating incorporated in G11. Supernovae feedback, based on \citet{martin1999properties}, is modeled in G11 as:
\begin{equation}
\delta M_{reheat} = \epsilon_{disc} \times \delta M_{\star}
\label{bh1}
\end{equation}
where $\delta M_{\star}$ is the stellar mass of the newly formed stars over some finite time interval. Unlike DLB07, G11 has a variable $\epsilon_{disk}$ and is modeled as follows:
\begin{equation}
\epsilon_{disk} = \epsilon \times \left[0.5 + \left(\frac{V_{max}}{70 kms^{-1}}\right)^{-\beta}\right]
\label{bh2}
\end{equation}
where both $\epsilon$ and $\beta$ are free parameters which were set in G11 based on the observed stellar mass function. 
\par 
In most SAMs, when a merger happens, all the hot gas is assumed to be transferred instantaneously from the smaller halo to the larger halo; however, this rapid transfer has been shown to cause a rapid decline in star formation \citep{baldry2006galaxy, wang2007modelling}. G11 implements a gas stripping model that includes both the instantaneous stripping and tidal, more gradual stripping in their treatment. 

\par	There are two reasons why we chose to use the Millennium simulation over more recent, higher-resolution simulations with a more accurate $\Lambda$CDM cosmogony. First, Millennium is a state of the art cosmological simulation that is uniquely linked to the concurrent development and refinement of two cutting edge SAMs \citep{croton2006many, bower2006breaking}. Second, and perhaps more important, the Millennium simulation provides a readily accessible data set. The publicly available simulation data  enables reproducibility and consistency. In the growing climate of scientific reproducibility, this approach is becoming increasingly important; and we follow this trend and release all our data and code at: \href{https://github.com/ProfessorBrunner/ml-sims}{\texttt{https://github.com/ProfessorBrunner/ml-sims}}. Next, we describe how the data set was obtained.

\subsection{Data Extraction} \label{data_extraction}

We used the online SQL database hosted by GAVO \citep{lemson2006halo} to construct our data set. Using the queryable SQL database for the Millennium Simulation, we extracted 365,361 dark matter haloes at $z=0$. Only dark matter haloes with masses larger than $10^{12} h^{-1}M_{\odot}$ were used in our analysis. For each dark matter halo, we extracted the following physical properties: number of dark matter particles ($\mathcal{N}$), spin, $M_{crit200}$, maximum circular velocity ($v_{max}$), and velocity dispersion ($\sigma_v$). For haloes at $z=0$, we also include the virial mass $M_{virial}$, the half-mass radius $R_{half}$, virial velocity $v_{virial}$, virial radius $r_{virial}$ and $r_{crit,200}$. Furthermore, we extracted the cold gas mass, total stellar mass, stellar mass in the bulge mass, central black hole mass and the hot gas mass of the primary galaxy for each dark matter halo and matched them likewise. As discussed in the introduction, the merger history of each dark matter halo plays an important role in how SAMs populate dark matter haloes with galaxies. However, sampling the merger history sufficiently and translating that into a well-defined set of inputs for our ML algorithms turns out to be a difficult task. 
\par	The naive way to extract the merger history would be to take all the progenitor haloes of each dark matter halo and use the internal properties of each progenitor halo and the current halo as the  inputs to our ML algorithms. However, in the Millennium simulations, some massive dark matter haloes have tens of progenitors just one snapshot back, potentially making the number of inputs in the hundreds. Moreover, going just one snapshot back is not sufficient to truly capture the merger history of a dark matter halo. Consequently, going to higher redshifts would easily result in the number of inputs being in the thousands. Our algorithms' runtimes are directly dependent on the dimension of the input data, and, therefore, employing the naive approach would severely impact the efficiency of our ML algorithms. 
\par	The merger tree for the Millennium simulation includes a descendantID and a firstProgenitorID for each halo. We are left with two ways to sample the merger history while retaining computational efficiency. We can either go top-down (i.e., start at a high redshift and go to $z=0$) by using descendantID or we could go bottom-up (i.e. starting at $z=0$ and proceed to higher redshifts) by using the firstProgenitorID. In the Millennium documentation\footnote{http://gavo.mpa-garching.mpg.de/Millennium/Help/mergertrees}, the first progenitor is defined as `the main progenitor of the subhalo'. The first progenitor is simply defined as the most massive of each dark matter halo's progenitors. And thus the firstProgenitorID tracks the main branch of a merger tree. We chose the latter approach for two reasons. The first approach results in fewer haloes and, consequently, galaxies, being examined at $z=0$. Second, the `main progenitor' approach encodes information about the main branch of a progenitor of each halo, implying that this may provide more valuable information about the dark matter halo's internal history. Using the firstProgenitorID from $z=0$ to $z=5.724$ (from snapshot 63 to 19), we extract the five physical properties mentioned above for 365,361 DM haloes. 
	\par Dark matter haloes below $10^{12} M_{\odot}$ weren't considered in our analyses because the computational cost associated with our technique (given how far back we go in the merger tree) would rise significantly since the number of haloes between just $10^{11}$ and $10^{12} M_{\odot}$ is a factor of O(100) to O(1000) greater than what we've considered in this work. This amounted to a tradeoff between a higher range of masses explored and how much deeper we can go into the merger tree. Since the main point of the paper is to roughly explore the applicability of ML in reproducing a reasonable population of galaxies, we decided against including haloes of lower masses. However, lower mass haloes are included in our analyses in the next paper \citep{kamdar2015machine} in the series where we apply machine learning to N-body + hydrodynamical simulations. We show there that when haloes of all masses are considered and we see that there is more scatter at lower masses but a reasonably high amount of information is recovered from the dark matter halo properties using machine learning throughout.
\subsection{Machine Learning} \label{ml}
In this subsection, we briefly talk about the basics of ML and outline the best performing algorithms.
\subsubsection{Overview} \label{ml_overview}
Machine learning is a bustling field in computer science, with a wide variety of applications in a number of other areas. The basic idea of ML algorithms is to `learn' relationships between the input data and the output data without any explicit analytical prescription being used. Supervised learning techniques are provided some training data $(\textbf{X, y})$ and they try to learn the mapping $G(\textbf{X} \rightarrow \textbf{y})$ in order to apply this mapping to the test data. 
\par Machine learning has been applied to several subfields in astronomy with a lot of success; see, for example, \citet{ball2010data, ivezic2014statistics} for a review of the applications of machine learning to astronomy. A decent majority of the applications of ML in astronomy have either been in classification problems such as star-galaxy classification \citep{ball2006robust, kim2015hybrid}, galaxy morphology classification \citep{banerji2010galaxy, dieleman2015rotation};  or have been regression applications like: photometric redshift estimation  \citep{ball2007robust, gerdes2010arborz, kind2013tpz}, and estimation of stellar atmospheric parameters \citep{fiorentin2007estimation}. 

\par To the best of our knowledge, however, only a few have applied ML to the problem of galaxy formation and the galaxy-halo connection. \citet{xu2013first}, which inspired this paper, predicted the number of galaxies in a dark matter halo to create mock galaxy catalogs. They used k-Nearest Neighbors (kNN) and Simple Vector Machines (SVM) to obtain promising results. Furthermore, \citet{ntampaka2015machine} used machine learning for dynamical mass measurements of galaxy clusters also showing promise. Given their non-parametric nature and incredibly powerful predictive capabilities, machine learning provides an attractive and intriguing method to study galaxy formation and evolution. 
\par For our study, we used a variety of machine learning algorithms: kNN, decision trees, random forests and extremely randomized trees. To quantify how well the algorithms are doing at learning relationships in the data, we use the mean squared error (MSE) metric. The MSE is defined as: 
\begin{equation}
MSE = \frac{1}{N_{test}}\sum_{i=1}^{i=N_{test}-1} \left(X_{test}^{i} - X_{predicted}^{i}\right)^2
\end{equation} 
Here, $X_{test}^i$ is the $i^{th}$ value of the actual test set and $X_{predicted}^i$ is the $i^{th}$ value of the predicted set. Furthermore, to gauge the relative performance of the algorithms, we also introduce the base MSE, following in the footsteps of \citet{xu2013first}, defined as:
\begin{equation}
MSE_b = \frac{1}{N_{test}}\sum_{i=1}^{i=N_{test}-1} \left(X_{test}^{i} - X_{mean,train}\right)^2
\end{equation}
Here, $X_{mean,train}$ is, as the name suggests, the mean of the training data set. $MSE_b$ is an extremely naive prediction of the error since each test point is simply predicted as the mean of the training dataset. $MSE_b$ will serve as an extremely useful metric when we want to measure the relative performance of our machine learning algorithms and the factor $\frac{MSE_b}{MSE}$ will quantitatively show how good our model is at minimizing error. \textit{The lower the MSE, and consequently the higher the factor $\frac{MSE_b}{MSE}$, imply a more robust prediction. }

\par 
Furthermore, we will also be using the following two metrics to check for the robustness of the prediction: the Pearson correlation and the coefficient of determination (`regression score function'). The pearson correlation is defined as 
\begin{equation}
\rho = \frac{cov(X_{predicted} X_{test})}{\sigma_{X_{predicted}} \sigma_{X_{test}}}
\end{equation}
 and the coefficient of determination is defined as: 
 \begin{equation}
 R^2 = 1 - \frac{\sum_i (X_{test}^i - X_{predicted}^i)^2}{\sum_i (X_{test}^i - X_{mean,train})^2}
 \end{equation}
 \textit{A higher $\rho$ and $R^2$ imply a robust prediction.} As shown in the results section, extremely randomized trees and random forests consistently outperform the other algorithms; consequently, we now briefly review the basics of these two techniques.

\begin{table*}
\begin{minipage}{140mm}
 \caption{An outline of the extremely randomized trees regression algorithm}
 \label{symbols}
 \begin{tabular}{@{}lcccccc}
  \hline
  \hline
  \begin{centering}
  Extremely Randomized Trees
  \end{centering} \\
  \hline
  \textbf{Inputs}: A training set $S$ corresponding to \textbf{$(X, y)$} input-output vectors, where \\ \textbf{X}=$(X_1,X_2,...,X_N)$ and  \textbf{y}=$(y_1,y_2,...,y_l)$, M (number of trees in the ensemble), \\ K (number of random splits screened at each node) and $n_{min,samples}$ (number of samples \\ required to split a node)\\
  \textbf{Outputs}: An ensemble of M trees: $\mathcal{T} = (t_1, t_2,...,t_M)$ \\
  \hline
  \textit{\textbf{Step 1}}: Randomly select $K$ inputs $(X_1,X_2,...,X_K)$ where $1\leq K \leq N)$. \\
  \hline 
  \textit{\textbf{Step 2}}: For each selected input variable $X_i$ in $i=(1,2,...,K)$: \\
  \tabitem Compute the minimal and maximal value of $X$ in the set: $X_i^{min}$ \\ and $X_i^{max}$ \\
  \tabitem Randomly select a cut-point $X_c$ in the interval [$X_i^{min}$, $X_i^{max}$] \\
  \tabitem Return the split in the interval $X_i \leq X_c$ \\
  \hline
  \textit{\textbf{Step 3}}: Select the best split $s_{\star}$ such that Score($s_{\star}$, S) = $max_{i=1,2,...,K}$ Score($s_{i}$, S) \\
  \hline
  \textbf{\textit{Step 4}}: Using $s_{\star}$, split $S$ into $S^l(X_i)$  and $S^r(X_i)$ \\
  \hline
  \textbf{\textit{Step 5}}: For $S^l(X_i)$  and $S^r(X_i)$, check the following conditions: \\
  \tabitem $|S^l(X_i)|$ or $|S^r(X_i)|$ is lower than $n_{min,samples}$ \\
  \tabitem All input attributes $(X_1,X_2,...,X_N)$ are constant in $|S^l(X_i)|$ or $|S^r(X_i)|$ \\
  \tabitem The output vector $(y_1,y_2,...,y_l)$ is constant in $|S^l(X_i)|$ or $|S^r(X_i)|$ \\
  \hline
  \textbf{\textit{Step 6}}: If any of the conditions in step 5 are satisfied, stop. We're at a leaf node. \\
  If none of the conditions are satisfied, repeat steps 1 through 5. \\
  \hline 
\end{tabular}
\end{minipage}
\end{table*}

\subsubsection{Extremely Randomized Trees} \label{etr}
Extremely randomized trees (ERT) is an ensemble learning technique that builds upon the widely used decision trees (for the purposes of regression, decision trees are called regression trees). Therefore, to understand how ERT works, we must first discuss the fundamentals of  regression trees. What follows is only a brief overview of both techniques; for a more comprehensive account of the technique, the reader is referred to \citet{breiman1984classification} and \citet{geurts2006extremely}. Regression trees follow a relatively simple procedure: \\[2ex]
\tabitem \textbf{\textit{Step 1:}} Construct a node containing all the data points and compute $m_c$ and $S$, where $m_c$ and $S$ are defined as:
\begin{equation}
m_c = \frac{1}{n_c} \sum_{i\epsilon c} z_i
\end{equation}
\begin{equation}
S(M) = \sum_{c \in leaves(M)} \sum_{i \in c} (z_i - z_c)^2
\label{tree_eqn}
\end{equation}
Here, $c$ are the possible values of dimension M, $z_i$ gives the target value on each branch $c$ and $z_c$ gives the mean value on that branch $c$. We can, therefore, rewrite Equation \ref{tree_eqn} as:
\begin{equation}
S(M) = \sum_{c\in leaves(M)} = n_c V_c
\end{equation}
$S(M)$ signifies the sum of the squared errors for some node M,\textit{ where $n_c$ is the number of samples in a leaf c and $V_c$ is the variance in leaf c.} \\
\tabitem \textbf{\textit{Step 2:}} If all the points in the node have the same value for all the input variables, we stop the algorithm. Otherwise, we  scan over all dimension splits of all variables to find the one
that will reduce S(M) as much as possible. If the largest decrease in S(M) is less than some threshold $\epsilon$, we stop the algorithm. Otherwise, we take that split, creating new nodes of the specified dimension. \\
\tabitem \textbf{\textit{Step 3:}} Go to Step 1 \\

\par Regression trees are usually considered to be weak learners. A technique that is used to turn these weak learners into strong ones involves building an ensemble of weak learners. In the context of regression trees, there are two popular ensemble methods: boosting and randomization. Since we have a multidimensional output, we focus on randomized ensemble techniques. The essence of ERT is to build an ensemble of regression trees where both the attribute and split-point choice are randomized while splitting a tree node. We provide pseudocode for the full algorithm in Table \ref{symbols}, which closely follows the algorithm outlined in \citet{geurts2006extremely}. In the algorithm, the Score is the relative reduction in the variance. For the two subtrees $S^l$  and $S^r$ corresponding to the split $s_{\star}$, the Score($s_{\star}$, S) (abbreviated to $Sc(s_{\star}, S)$) is given by:
\begin{equation}
Sc(s_{\star}, S) = \frac{var(\textbf{y}, S) - \frac{|S^l|}{|S|} var(\textbf{y}, S^l) - \frac{|S^r|}{|S|} var(\textbf{y}, S^r)}{var(\textbf{y},S)}
\end{equation}

The estimates produced by the $M$ trees in the ERT ensemble are finally combined by averaging $y$ over all trees in the ensemble. The use of the original training data set in place of a bootstrap sample (as is done for random forests) is done to minimize bias in the prediction. Furthermore, the use of both randomization and averaging is aimed at reducing the variance of the prediction \citep{geurts2006extremely}. 

\subsubsection{Random Forests} \label{rf}

The methodology of random forests \citep[RF]{breiman2001random} is very similar to that of extremely randomized trees. There are two central algorithmic differences between the two methods. First, RF use a bootstrap replica \citep{breiman1996bagging}. Bootstrap replica consists of selecting a random sample without replacement from the training data ($\textbf{X,y}$). ERT, on the other hand, uses the original training data set. Second, ERT chooses the split randomly from the range of values in the sample at each split, whereas RF tries to determine the best split at each internal node. We briefly outline the basic steps of the algorithm in Table \ref{rf}. 

\begin{table}
 \caption{An outline of the random forests algorithm}
 \label{rf}
 \begin{tabular}{@{}lcccccc}
  \hline
  \hline
  Random forests \\
  \hline
  \textbf{Inputs}: A training set $S$ corresponding to \textbf{$(X, y)$} input-output \\ vectors, where \textbf{X}=$(X_1,X_2,...,X_N)$ and  \textbf{y}=$(y_1,y_2,...,y_l)$, \\ M (number of trees in the ensemble), \\  $K$ (number of features to consider when looking for best split) \\ $n_{min}$ (minimum number of samples required to split a node)\\
  \textbf{Outputs}: An ensemble of M trees: $\mathcal{T} = (t_1, t_2,...,t_M)$ \\
  \hline
  \textbf{\textit{Step 1}}: Select a new bootstrap sample from the training data \\ set \\
  \hline
  \textbf{\textit{Step 2}}: Grow an unpruned tree on this bootstrap \\
  \hline 
  \textbf{\textit{Step 3}}: For each node in the tree, randomly select $K$ features, \\ look for the best split using only these features \\
  \hline
  \textbf{\textit{Step 4}}: Save the tree and do not perform pruning. \\ 
  \hline
  \textbf{\textit{Step 5}}: Perform steps 1 through 4 $M$ times. \\
  \hline
  \textbf{\textit{Step 6}}: The overall prediction is the average output from each \\ tree in the ensemble. \\
	\hline
\end{tabular}
\end{table}

\par For our analyses, we used the implementation of ERT and RF provided in the Python library scikit-learn \citep{pedregosa2011scikit}. The parameters we used and the runtime of the techniques for the problem are discussed in the results section. ERT tends to be faster than RF because of the randomization in finding the split, reducing the training time. The reduced training time lets us build a bigger ensemble of trees and explains why our ERT results are, generally, marginally better than RF's. 

\section{Results} \label{results}
In this section, we present and discuss the results that were obtained when we applied the algorithms to the Millennium data. Using the dark matter internal halo properties and a partial merger history as our inputs, the following components of the final mass of the galaxy are predicted: stellar mass in the bulge, total stellar mass, cold gas mass, central black hole mass and hot gas mass. In nature, these attributes are the result of billions of years of dissipative, nonlinear baryonic processes. As discussed earlier, the basic ingredient for large scale structure formation is the $\Lambda$CDM model; but, on a smaller scale, the story is incredibly different and vastly more complicated. In this section, we try to draw a link between the two regimes using ML algorithms to explore the halo-galaxy connection. We first discuss the performance of ML in reproducing the simulated properties of the galaxies in G11 and the implications of our results for the halo-galaxy connection. Then, we address some discrepancies in our results (particularly the cold gas mass), discuss why the cold gas mass prediction is not robust and provide an alternative, significantly more accurate model that includes two baryonic inputs over two snapshots.

\par
Table \ref{mse} lists the results we obtained with the different machine learning algorithms for each component of the mass of the central galaxy. $MSE_b$ and the $MSE$ are listed for each technique. The factor reduction of the $MSE$ is also listed to test the relative performance of the algorithms to quantify how much they are learning. Finally, the pearson correlation between the predicted and the true data set and the coefficient of determination ($R^2$) are also listed. Seventeen different plots (Figures 2 through 18) are included to show the best results we obtained by using both ERT and RF. A hexbin plot and a violinplot are shown for each component of the mass to compare the predictions to the G11 test data. \textit{A hexbin plot is a 2D histogram and provides information about the goodness of fit. A kernel density estimate (KDE) is plotted on each axis to overlay information about the distribution as well. A violinplot is a boxplot with a KDE on the side, providing more information about the distribution of a particular set.} Furthermore, a stellar mass-halo mass relation plot is  shown to compare the physical reasonability of our stellar mass results with G11 results. A plot showing the cold gas mass fraction as a function of stellar mass is also included. Lastly, the G11 and ML BH mass-bulge mass relation for the predicted galaxies and G11 galaxies is shown in Figure \ref{bhbulge}. All plots were created using Seaborn\footnote{http://stanford.edu/~mwaskom/software/seaborn/} and Matplotlib. For the hexbin plot, a gridsize of 30 was used and the colormap was logarithmically scaled. For the kernel density estimate (KDE) in the violinplot, the bandwidth was chosen using the \citet{silverman1986density} method and the density is evaluated on 1000 grid points. The violinplots serves two purposes: providing an insightful look into how good ML is at reproducing a similar population of galaxies as found in G11 and providing a zoomed in alternative of what the mass distribution looks like.

\begin{figure}
  \includegraphics[width=84mm]{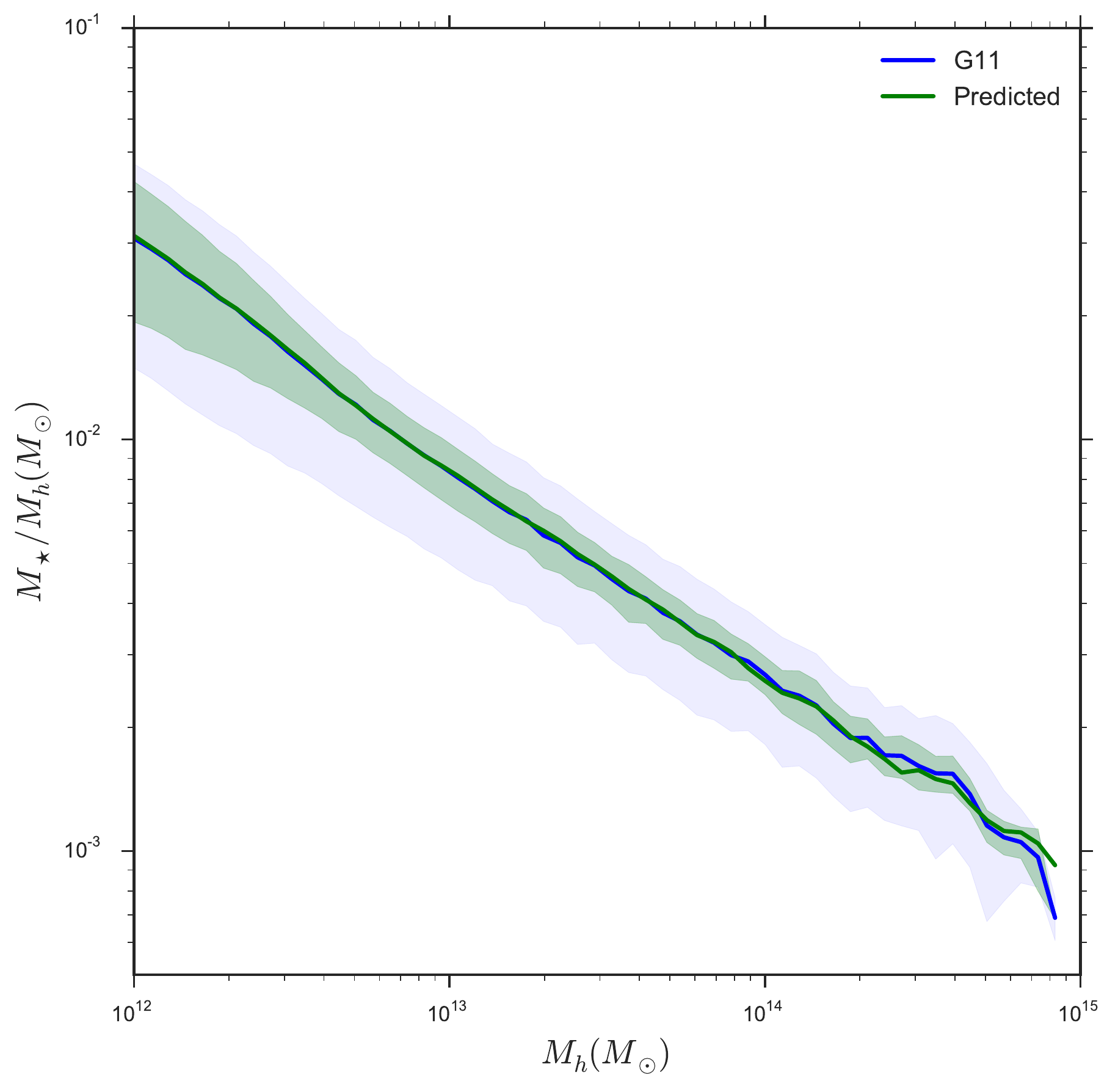}
  \caption{The stellar mass-halo mass relation for the predicted total stellar mass using machine learning and the total stellar mass in G11 are compared for central galaxies. Both quantities are binned using the halo mass from Millennium. The two different shadings (blue for G11 and green for ML) represent the standard deviation at each binned point for the respective technique.}
   \label{fig:smhm}
\end{figure}

\begin{figure}
\includegraphics[width=84mm]{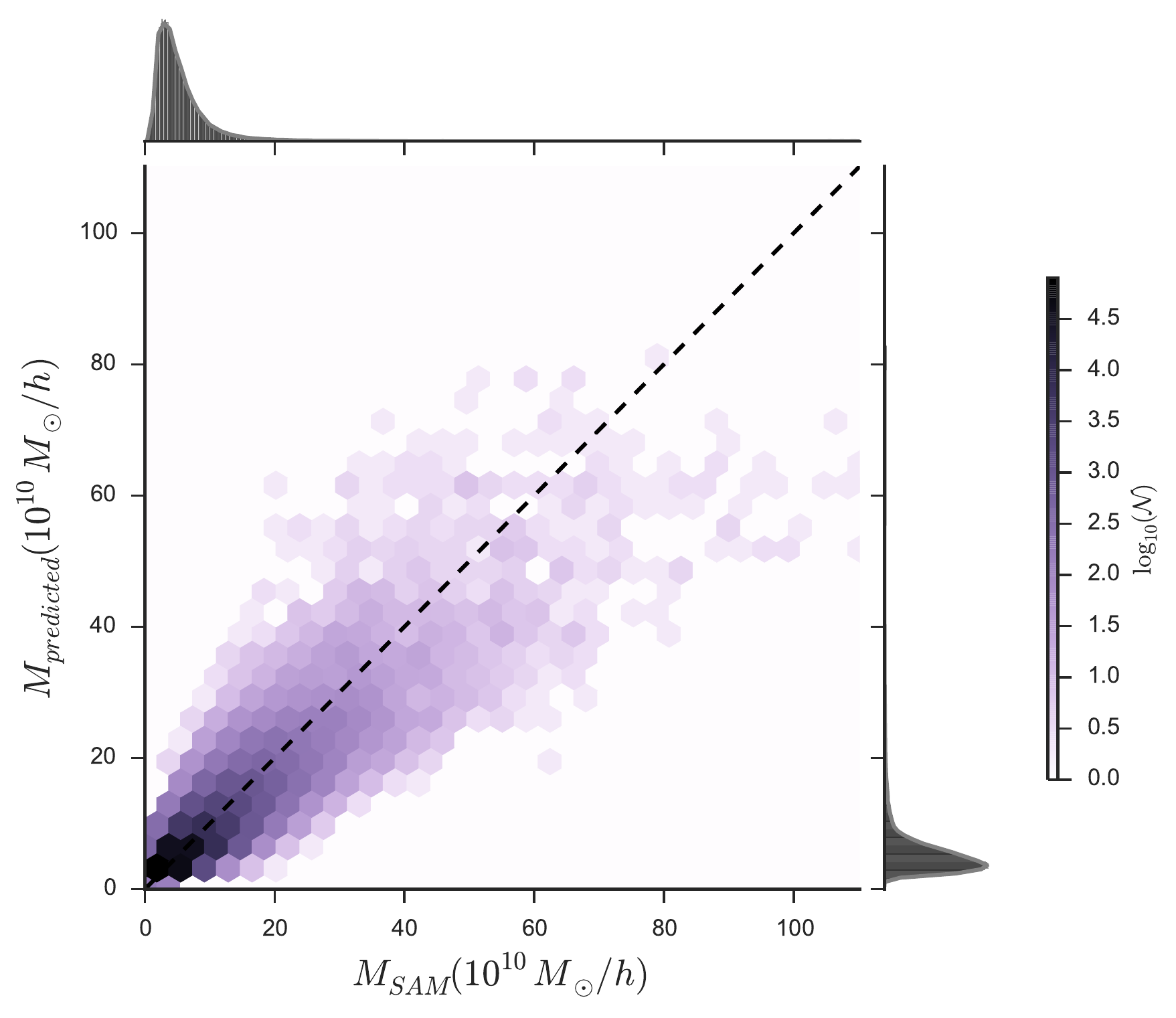}

\caption{A hexbin plot of $M_{SAM,\star}$ and $M_{predicted,\star}$. The black dashed line corresponds to a perfect prediction. The MSE for the prediction is 5.755, the Pearson correlation between the predicted galaxy mass and the G11 galaxy mass is 0.876 and the regression score is 0.768.}
\label{stellar1}
\end{figure}

\begin{figure}
  \includegraphics[width=84mm]{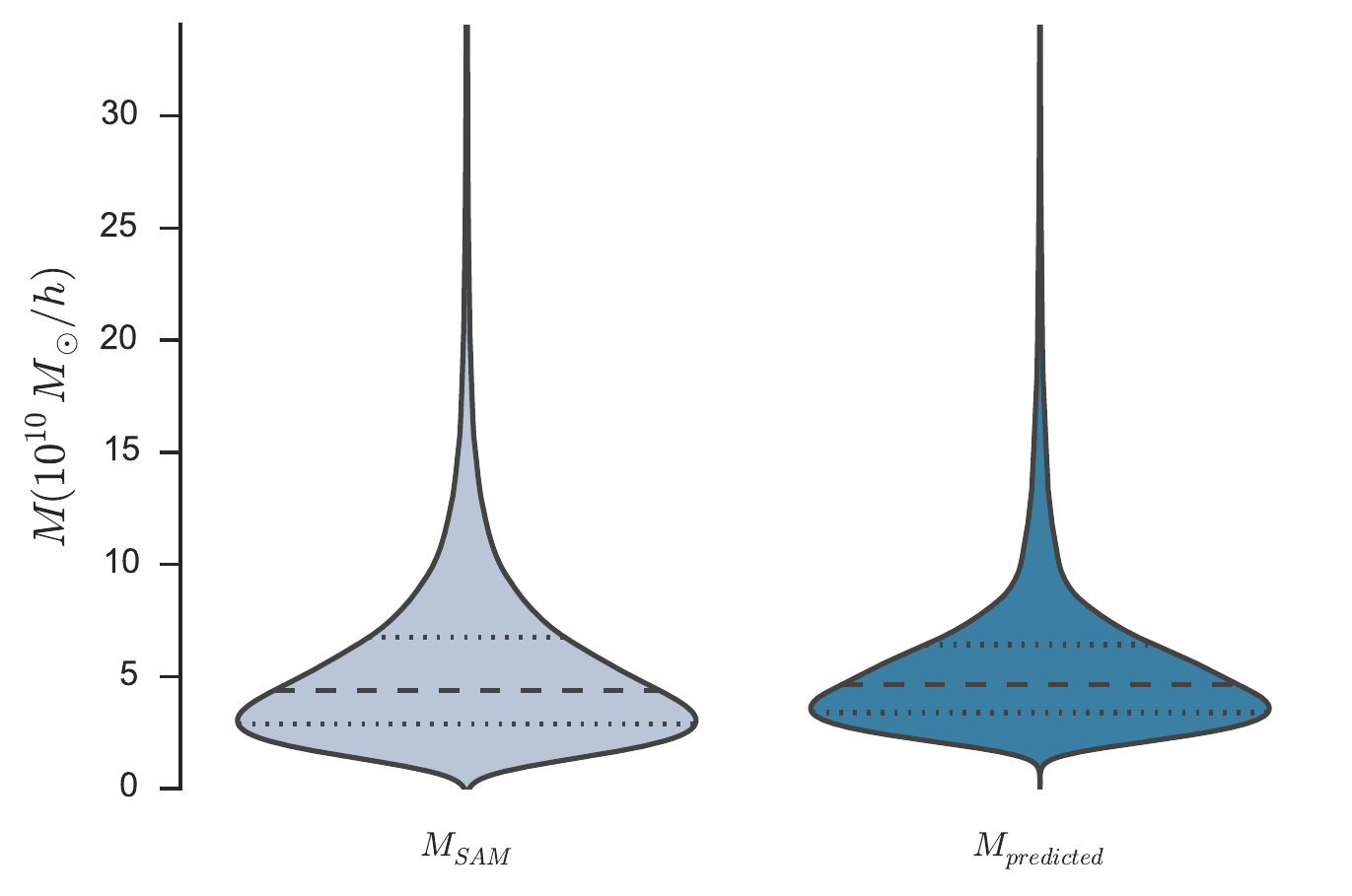}
  \caption{A violinplot is plotted that shows the distributions of the $M_{SAM,\star}$ and $M_{predicted,\star}$. The median and the interquantile range are shown for both sets of galaxy masses.}
    \label{stellar2}

\end{figure}

\begin{table*}
\begin{minipage}{159mm}
 \caption{The performance of different machine learning techniques in predicting the different mass components the central galaxy in each dark matter halo at $z=0$.}
 \label{mse}
 \begin{tabular}{@{}lcccccc}
  \hline
  \hline
   & Technique & $MSE_b$ & $MSE$ & Factor ($\frac{MSE_b}{MSE}$) & Pearson Correlation
         & $R^2$\\
  \hline
  & kNN & & 6.661 & 3.624 & 0.852 & 0.724 \\
  & Decision Trees &  & 7.448 & 3.241 & 0.832 & 0.691\\
  & Random Forests &  & 5.763 & 4.301 & \textcolor{darkgreen}{\textbf{\textit{0.876}}} & \textcolor{darkgreen}{\textbf{\textit{0.768}}}\\
  \multirow{-5}{*}{$M_{\star,total}$} & \textcolor{darkgreen}{\textbf{Extremely Randomized Trees}} & \multirow{-5}{*}{24.788} & \textcolor{darkgreen}{\textit{\textbf{5.755}}} & \textcolor{darkgreen}{\textbf{\textit{4.307}}} & \textcolor{darkgreen}{\textbf{\textit{0.876}}} & 0.766\\
  
  \hline
  & kNN & & 7.066 & 3.906 & 0.864 & 0.744 \\
  & Decision Trees &  & 8.012 & 3.444 & 0.843 & 0.710 \\
  & \textcolor{darkgreen}{\textbf{Random Forests}} &  & \textcolor{darkgreen}{\textbf{\textit{6.305}}} & \textcolor{darkgreen}{\textbf{\textit{4.467}}} & \textcolor{darkgreen}{\textbf{\textit{0.881}}} & \textcolor{darkgreen}{\textbf{\textit{0.775}}} \\
  \multirow{-5}{*}{$M_{\star,bulge}$} & Extremely Randomized Trees & \multirow{-5}{*}{28.165} & 6.306 & 4.466 & \textcolor{darkgreen}{\textbf{\textit{0.881}}} & \textcolor{darkgreen}{\textbf{\textit{0.776}}}\\
  \hline
  & kNN & & 1243.182 & 47.786 & 0.991 & 0.979\\
  & Decision Trees &  & 144.537 & 411.014 & \textcolor{darkgreen}{\textbf{\textit{0.999}}} & 0.998\\
  & Random Forests &  & 70.398 & 929.358 & \textcolor{darkgreen}{\textbf{\textit{0.999}}} & \textcolor{darkgreen}{\textbf{\textit{0.999}}} \\
  \multirow{-5}{*}{$M_{hot}$} & \textcolor{darkgreen}{\textbf{Extremely Randomized Trees}} & \multirow{-5}{*}{65424.910} & \textcolor{darkgreen}{\textbf{\textit{57.536}}} & \textcolor{darkgreen}{\textbf{\textit{1137.113}}} & \textcolor{darkgreen}{\textbf{\textit{0.999}}} & \textcolor{darkgreen}{\textbf{\textit{0.999}}}\\
 \hline
  & kNN & & 0.401 & 1.311 & 0.487 & 0.237\\
  & Decision Trees &  & 0.445 & 1.182 & 0.393 & 0.155\\
  & Random Forests &  & \textcolor{darkgreen}{\textbf{\textit{0.319}}} & 1.652 & \textcolor{darkgreen}{\textbf{\textit{0.631}}} & \textcolor{darkgreen}{\textbf{\textit{0.395}}} \\
  
  \multirow{-5}{*}{$M_{cold}$} & \textcolor{darkgreen}{\textbf{Extremely Randomized Trees}} & \multirow{-5}{*}{0.527} & \textcolor{darkgreen}{\textbf{\textit{0.319}}} & \textcolor{darkgreen}{\textbf{\textit{1.654}}} & \textcolor{darkgreen}{\textbf{\textit{0.632}}} & \textcolor{darkgreen}{\textbf{\textit{0.395}}} \\
 \hline
 & kNN & & 0.000063 & 6.958 & 0.926 & 0.856 \\
  & Decision Trees &  & 0.000081 & 5.432 & 0.903 & 0.815 \\
  & Random Forests &  & 0.000068 & 6.456 & 0.925 & 0.856 \\
  \multirow{-5}{*}{$M_{BH}$} & \textcolor{darkgreen}{\textbf{Extremely Randomized Trees}} & \multirow{-5}{*}{0.000439} & \textcolor{darkgreen}{\textbf{\textit{0.000066}}} & \textcolor{darkgreen}{\textbf{\textit{6.652}}} & \textcolor{darkgreen}{\textbf{\textit{0.927}}} & \textcolor{darkgreen}{\textbf{\textit{0.859}}}\\
 \hline
  
 \end{tabular}
 \end{minipage}
\end{table*}

\par 
The algorithms that performed the best, ERT and RF, were outlined in section 2. Using scikit-learn's implementation, we used the following parameters for ERT: $n_{trees} = 750$, and minimum sample split ($n_{min}$) = 5. For RF, we used the following parameters: $n_{trees} = 325$, and minimum sample split ($n_{min}$) = 5. We used 35\% of the G11 and Millennium data for training and the rest were used for testing. The entire pipeline for extremely randomized trees (includes data preprocessing, training, testing and generating all the plots) using the listed parameters ran on 2.7 GHz Intel Dual-Core Processor in 73 minutes. Likewise, the entire pipeline for random forests using the parameters above ran on the same system in 122 minutes. Note that in both cases, these times are orders of magnitude smaller than SAM computation times.

\begin{figure}
\includegraphics[width=84mm]{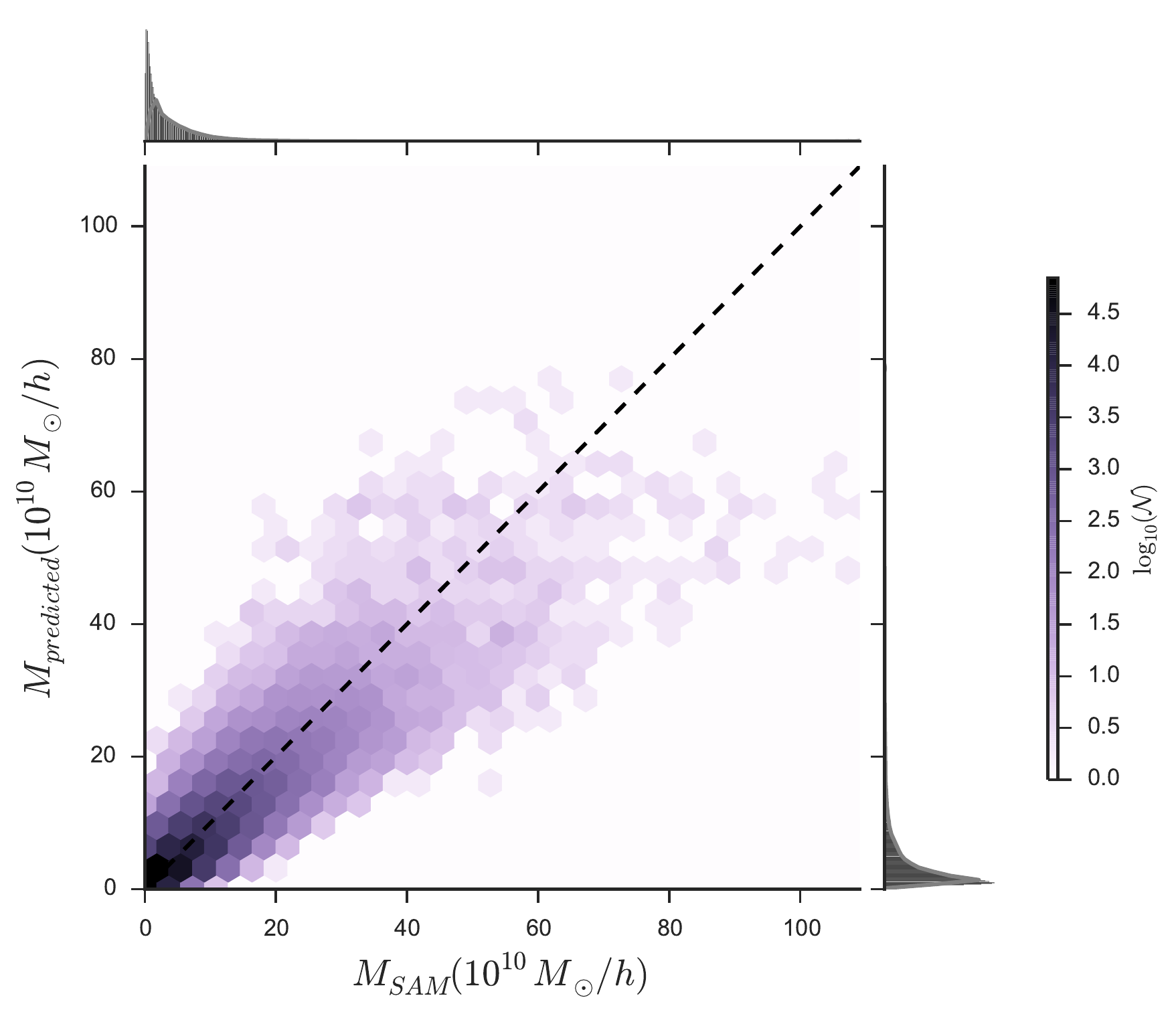}
\caption{A hexbin plot of $M_{SAM,bulge}$ and $M_{predicted,bulge}$. The black dashed line corresponds to a perfect prediction. The MSE for the prediction is 6.305, the Pearson correlation between the predicted galaxy mass and the G11 galaxy mass is 0.881 and the regression score is 0.775.}
\label{bulge1}
\end{figure}

\begin{figure}
  \includegraphics[width=84mm]{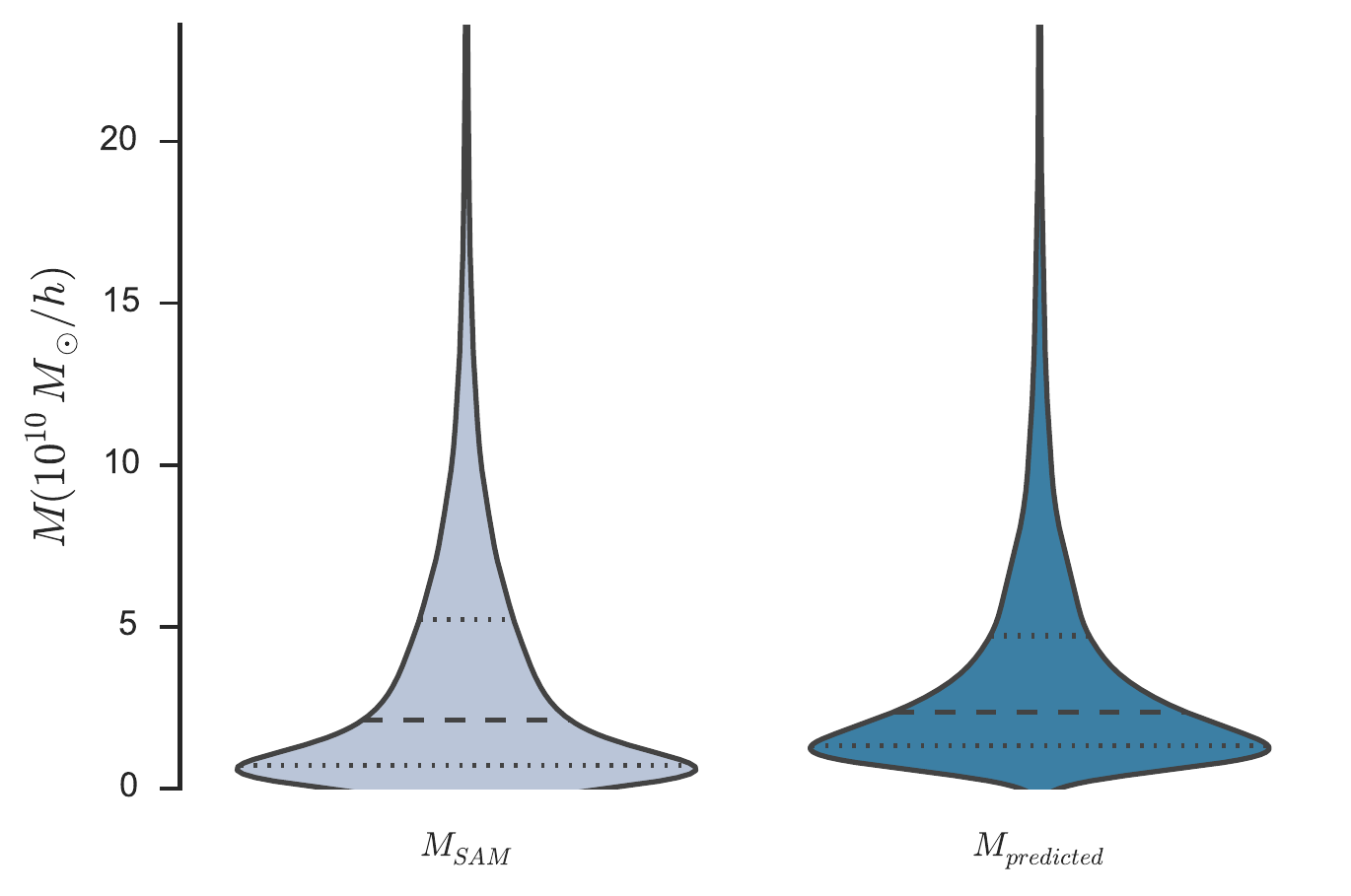}
  \caption{A violinplot showing the distributions of the $M_{SAM,bulge}$ and $M_{predicted,bulge}$. The median and the interquartile range for both sets of galaxy masses are shown. }
\label{bulge2}
\end{figure}

\begin{figure}
\includegraphics[width=84mm]{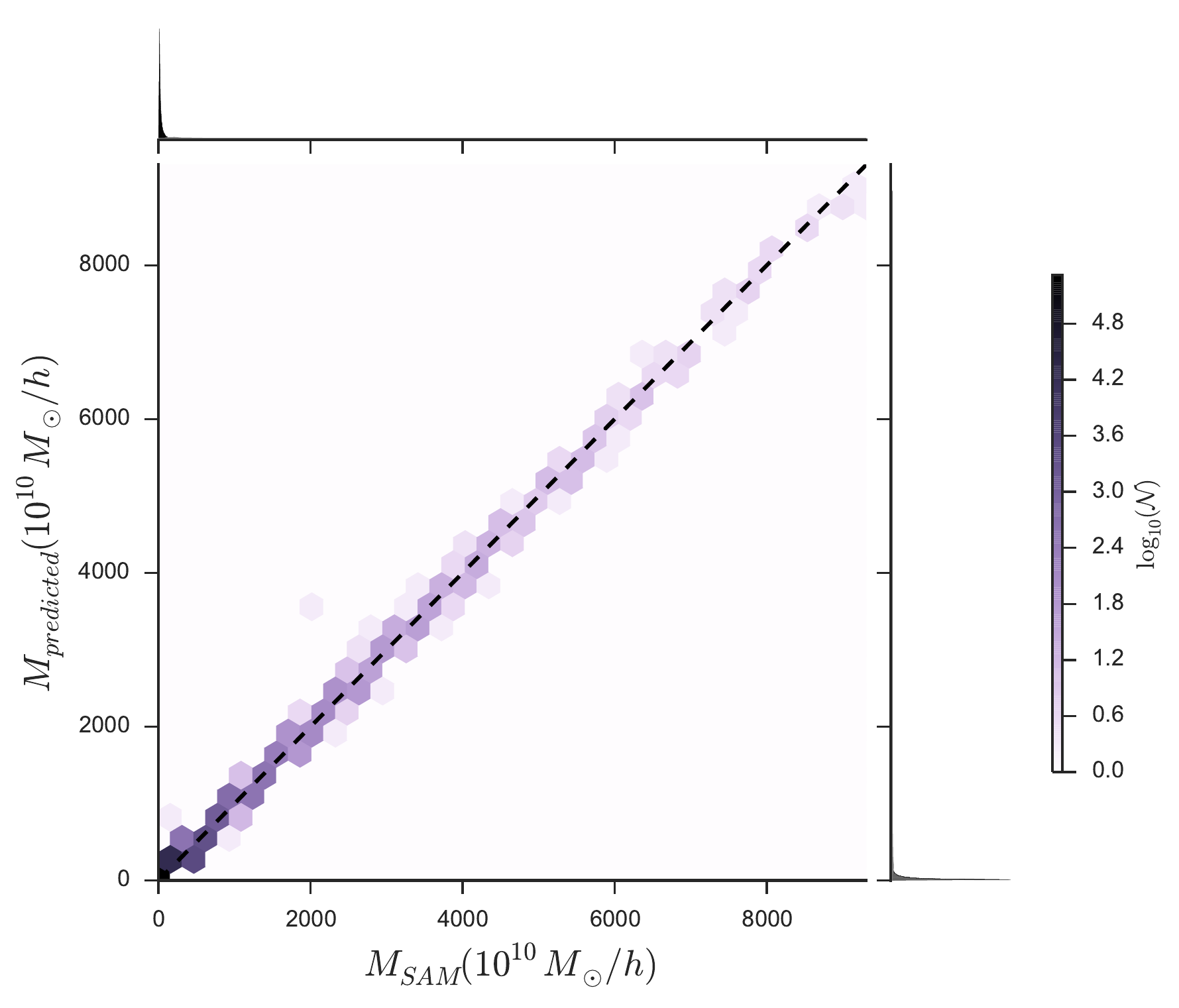}
\caption{A hexbin plot of $M_{SAM,hot}$ and $M_{predicted,hot}$. The black dashed line corresponds to a perfect prediction. The MSE for the prediction is 57.536, the Pearson correlation between the predicted galaxy mass and the G11 galaxy mass is 0.999 and the regression score is 0.999.}
\label{hot1}
\end{figure}

\begin{figure}
  \includegraphics[width=84mm]{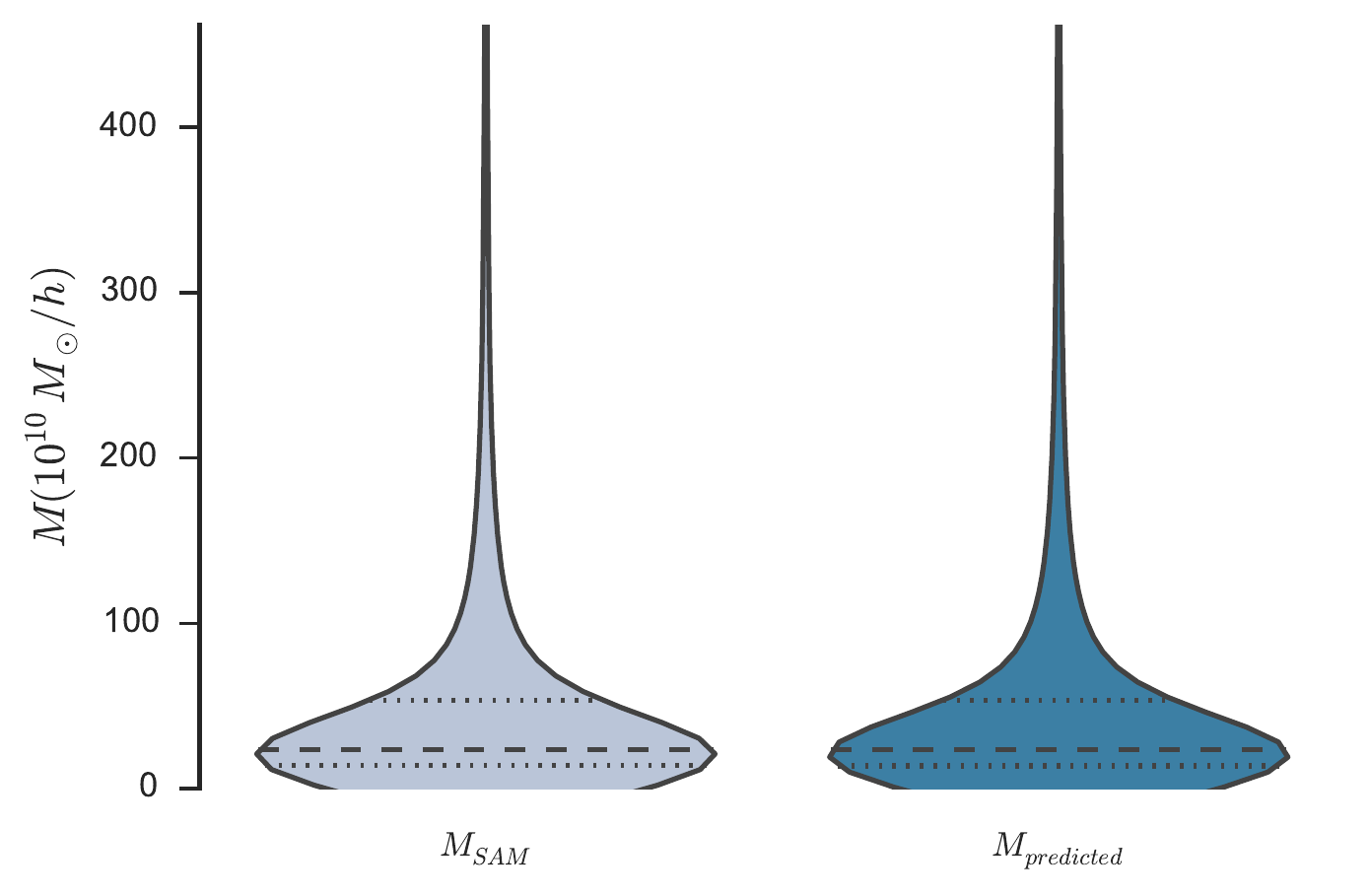}
  \caption{A violinplot showing the distributions of the $M_{SAM,hot}$ and $M_{predicted,hot}$. The median and the interquartile range for both sets of galaxy masses are shown.}
    \label{hot2}

\end{figure}

\begin{figure}
  \includegraphics[width=84mm]{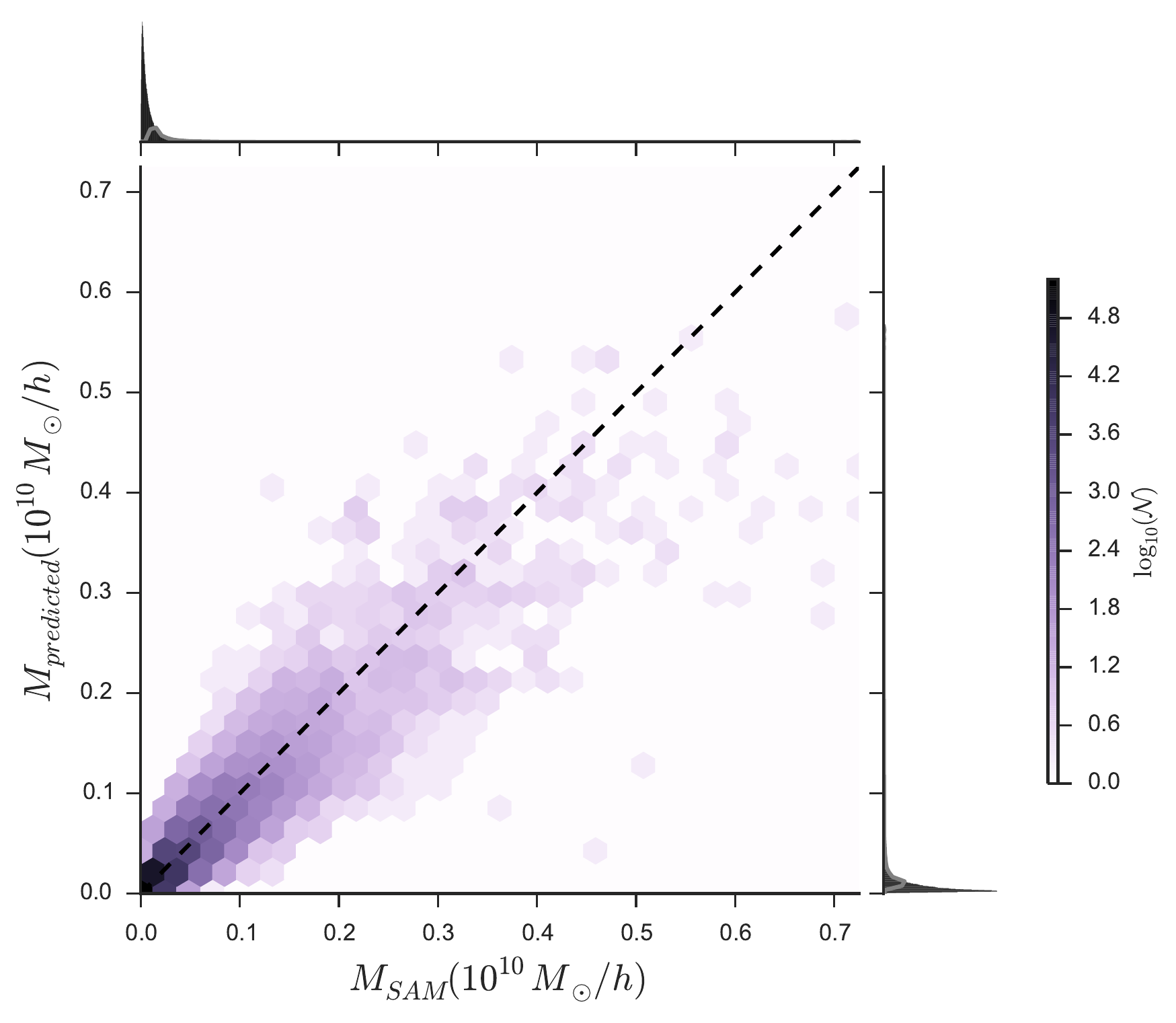}
  \caption{A hexbin plot of $M_{SAM,BH}$ and $M_{predicted,BH}$. The black dashed line corresponds to a perfect prediction. The MSE for the prediction is 0.000066, the Pearson correlation between the predicted galaxy mass and the G11 galaxy mass is 0.927 and the regression score is 0.86.}
    \label{black1}

\end{figure}

\begin{figure}
  \includegraphics[width=84mm]{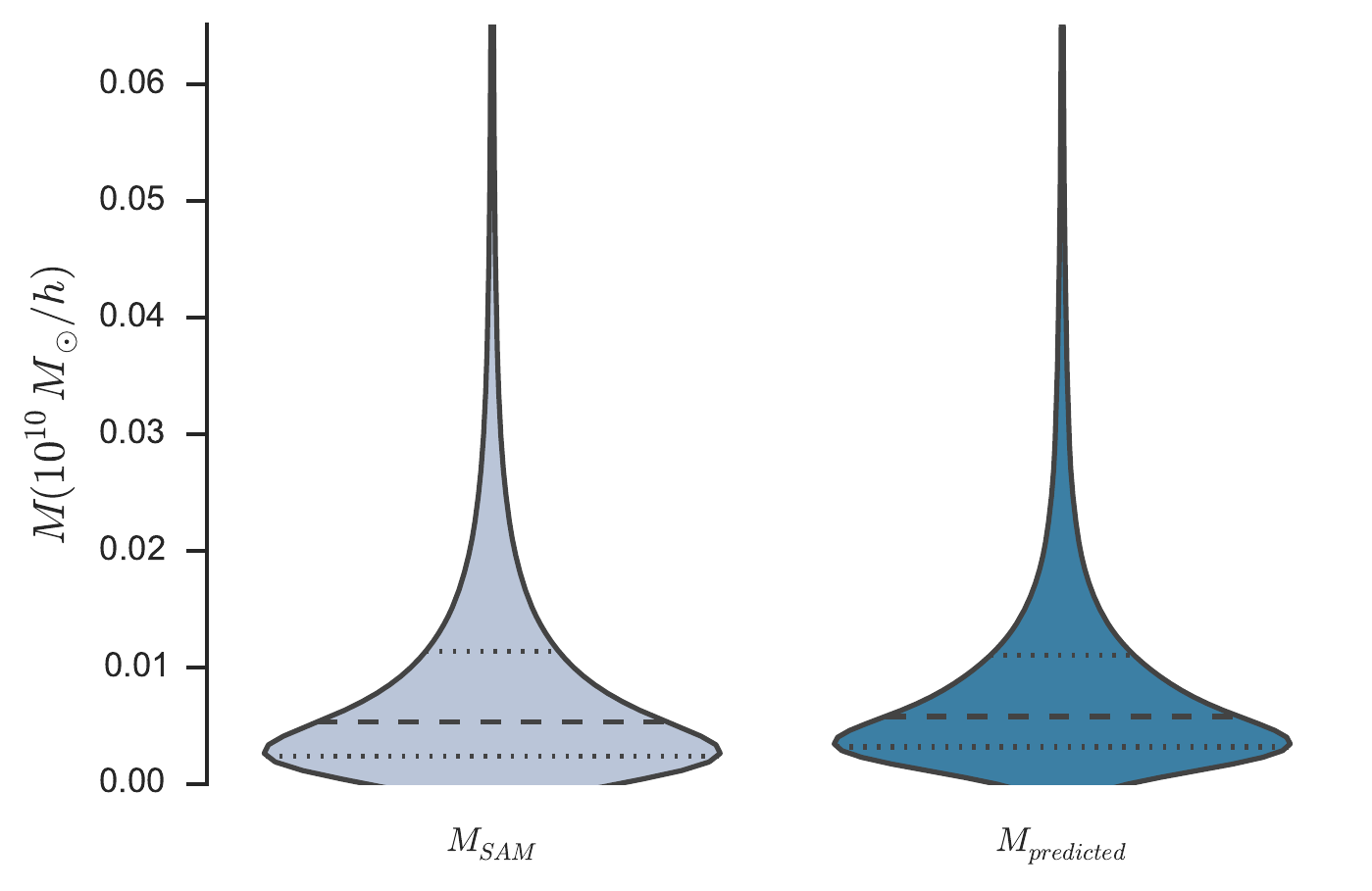}
  \caption{A violinplot showing the distributions of $M_{SAM,BH}$ and $M_{predicted,BH}$. The median and the interquartile range for both sets of galaxy masses are shown.}
    \label{black2}

\end{figure}

\begin{figure}
  \includegraphics[width=84mm]{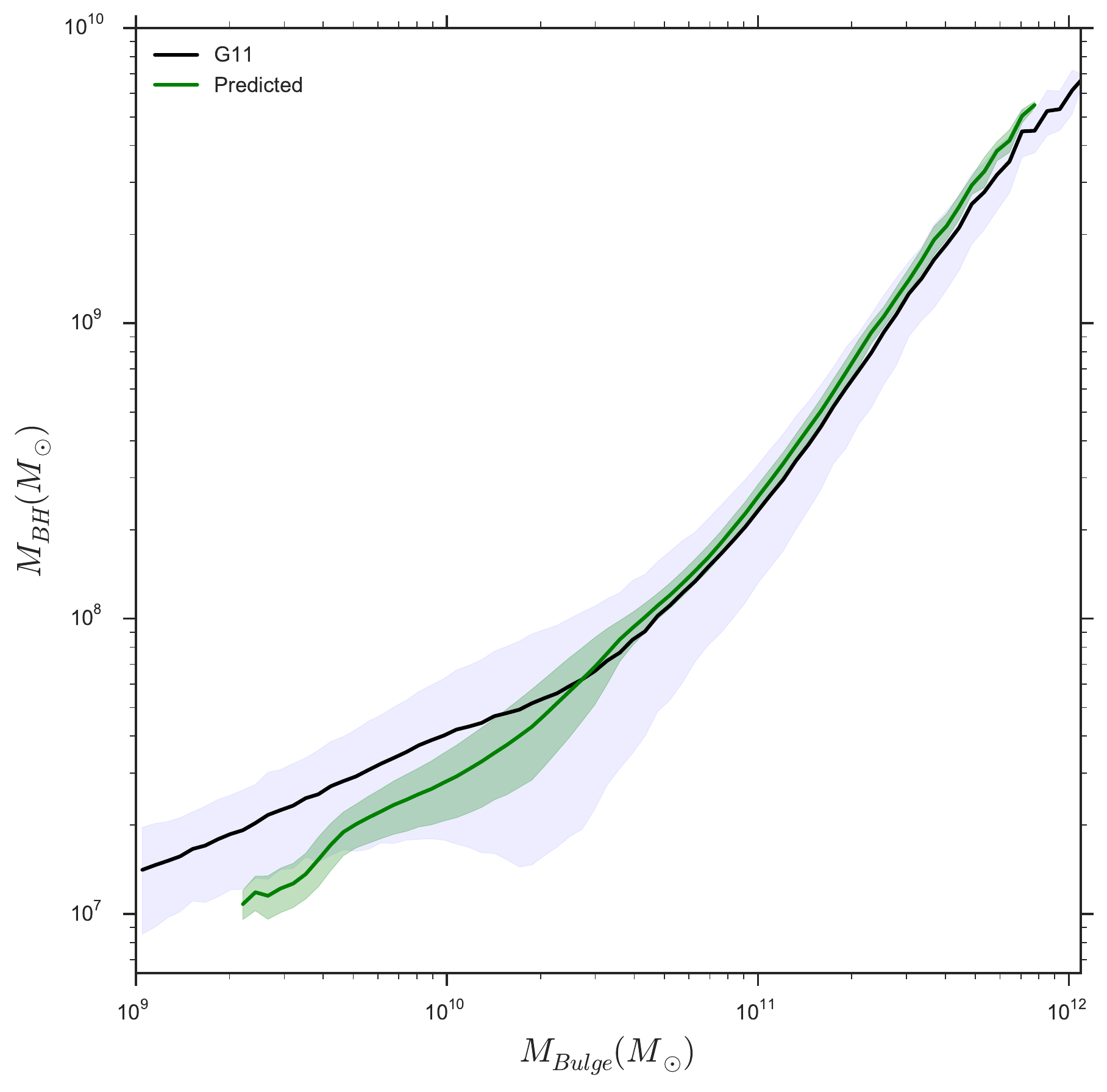}
  \caption{The black hole mass-bulge mass relation is plotted on a log scale for the predicted population of galaxies and G11's population of galaxies. For the predicted curve, $M_{BH,predicted}$ points are binned by the predicted bulge mass and for G11, $M_{BH,SAM}$ is binned by the corresponding G11 bulge mass. The shaded areas correspond to the standard deviation at each binned point (blue for G11 and green for ML).}
    \label{bhbulge}

\end{figure}

\subsection{Stellar Mass} \label{results_stellar}
\par
The first thing to notice in our results is the total stellar mass-halo mass relation (SMHM). In \citet{behroozi2010comprehensive} and \citet{moster2010constraints}, the SMHM is extensively studied and compared to a variety of observational data and prevalent empirical halo-galaxy models. The SMHM provides a very powerful tool to check whether our results seem physically meaningful and not just numerically reasonable. We can see in Figure \ref{fig:smhm} that the SMHM is reconstructed almost perfectly. The curves for the predicted set and the G11 results line up almost perfectly. One thing to notice here is that our prediction is slightly off for the higher halo masses. Moreover, there is more noticeable scatter in the Millennium SMHM than the reconstructed SMHM. These discrepancies are probably present because the ML algorithms are unable to model extreme cases, a hypothesis which is supported by the hexbin plot in Figure \ref{stellar1} as well; the galaxies with higher masses ($M_g > 60 \times 10^{10} M_{\odot}$) are being underpredicted. The SMHM being reproduced strongly implies that machine learning is able to approximate the mapping between the stellar mass and the halo mass that is prescribed in G11 very well. A subtlety to note here is that ML does not a priori assume that a direct mapping exists between the stellar mass and halo mass like other SMHM studies; instead, ML is trying to learn the relationship prescribed in G11 for how the galaxies are populated with stellar mass. This point will be important later in the paper when we compare our model with subhalo abundance matching. 
	
\par
Furthermore, we can clearly see in Figure \ref{stellar1} that the stellar mass is being predicted fairly well. The regression score ($R^2$) is $0.77$ and the correlation between the predicted set and the test G11 set is $0.876$. We can see clearly that, while there is some noticeable scatter, a fair majority of the predictions lie on the dashed black line implying that our predictions line up with that of G11's fairly well. Moreover, and perhaps more surprisingly, we can see in Figure \ref{stellar2} that the distribution of the stellar mass is reproduced perfectly using machine learning. Our model is able to pick up on the physical prescriptions that are used in G11 to populate galaxies with stars very well.

\subsubsection{Bulge Mass} \label{results_bulge}
\par 
We also predict the stellar mass that is in the bulge of each central galaxy at $z=0$. In Figures \ref{bulge1} and \ref{bulge2}, we can see that the bulge mass is also being accurately reproduced. The regression score is $0.775$ and the correlation between the predicted set is $0.881$. The hexbin plot shows that the bulge mass prediction is appreciably robust and very similar to the total stellar mass prediction. While the violinplot shows that the distribution of the bulge mass is also reproduced very well. The bulge mass is accumulated in G11 through two primary mechanisms: mergers and disc instabilities. The reproduction of the distribution of the bulge mass implies that both these physical prescriptions in G11 are being approximately well modeled by ML. 

\par However, there is a small discrepancy between the distribution of the true and the predicted set (Figure \ref{bulge2}); the predicted distribution shows that ML is overpredicting at lower masses. A possible explanation for this discrepancy is an overprediction of the stellar mass accumulated in the bulge in the event of mergers. We input the merger history of each dark matter halo into our model and not the galaxy-galaxy merger history. As outlined in \citet{hopkins2010mergers}, there are two ways to link the two regimes: defining a delay time scale, as discussed in \citet{boylan2008dynamical}, or tracking the subhaloes directly (only feasible for higher resolution simulations). Our robust predictions imply that ML is able to extrapolate a reasonable amount of information about galaxy-galaxy mergers through a halo-only merger history. However, the overprediction could easily be explained by ML's inability to fully pick up on the galaxy-galaxy merger timescale by using only a halo merger history. This idea is further supported by our prediction for the central black hole mass discussed in Section \ref{results_bh}. Overall, our model is able to pick up on the physical prescriptions that are used in G11 to model bulge formation with a minor discrepancy that has no trivial solution in the framework of our model.

\subsection{Hot Gas Mass} \label{results_hot}
The hot gas mass prediction, as shown in Figures \ref{hot1} and \ref{hot2} and Table \ref{mse}, is outstanding. The Pearson correlation is 0.999 and the regression score is 0.999. ML is able to pick up on the way that hot gas is modeled in G11 incredibly well. As discussed in section \ref{mdf: db_sam_hot}, the amount of hot gas available at each snapshot is directly dependent upon the total virial mass in the dark matter halo. Even though supernovae feedback plays an important role in reheating some of the gas found in the halo, ML is still able to pick up on how the prescriptions for the hot gas mass in G11 are set. The distribution of the hot gas mass is reproduced perfectly and the MSE is reduced by a factor of 1137. 

\par 
As discussed in section \ref{mdf: db_sam_hot}, the main contributors to the hot gas mass in the central galaxy are gas stripping and supernovae feedback. Our almost perfect results show that ML is able to model these two physical prescriptions very well. The former involves hot gas being stripped from a satellite galaxy and being added to the central galaxy for cooling. The latter, which plays a larger role in the determination of the hot gas mass, involves the reheating of cold gas due to supernova feedback. As outlined in Equations \ref{bh1} and \ref{bh2}, the amount of mass that is reheated has a partial dependence, both directly (in $\epsilon_{disk}$) and indirectly (in $\delta M_{\star}$, on halo properties and ML is able to model both quite well. The amount of hot gas mass plays an important role in the cooling (Equation \ref{cool}). The almost perfect predictions for the hot gas mass are promising and show ML's strength in modeling a mapping that is dominated by a direct analytical relationship (Equation \ref{hot_analytic}). 

\subsection{Central Black Hole Mass} \label{results_bh}
As discussed in Section \ref{mdf: db_sam_bh}, the central black hole mass is mostly accreted through the `quasar' mode (Equation \ref{quasar}) during major mergers or gas-rich mergers. Our results for the central black hole mass are shown in Table \ref{mse} and Figures \ref{black1} and \ref{black2}. The regression score is $0.86$ and the correlation is $0.929$, implying that our predictions match up well with the G11 data. An interesting point here is the overprediction of the black hole mass at lower masses shown in Figure \ref{black2}. This overprediction places confidence in the theory that the ML may be only be partially able to pick up on the galaxy-galaxy merger timescale that was outlined in the bulge mass discussion. 

\par In Figure \ref{bhbulge}, we show the predicted and the G11 BH mass-bulge mass relation. At higher bulge masses, the predicted and the G11 curves agree very well. For lower masses, there is a discrepancy between the bulge masses at which the curve starts. The predicted curve starts at a noticeably higher mass because of the overprediction of the bulge mass that was discussed earlier (i.e. the curve starts at the bin that corresponds to the lowest predicted bulge mass). Overall, the prediction for the central black hole mass and the reproduction of the BH-bulge relation reinforce the solidity of ML's predictive power in the context of galaxy evolution. Furthermore, the robust BH mass predictions imply that the internal dark matter halo properties and the merger history contain sufficient information to make robust predictions of the central black hole mass. 

\begin{figure}
\includegraphics[width=84mm]{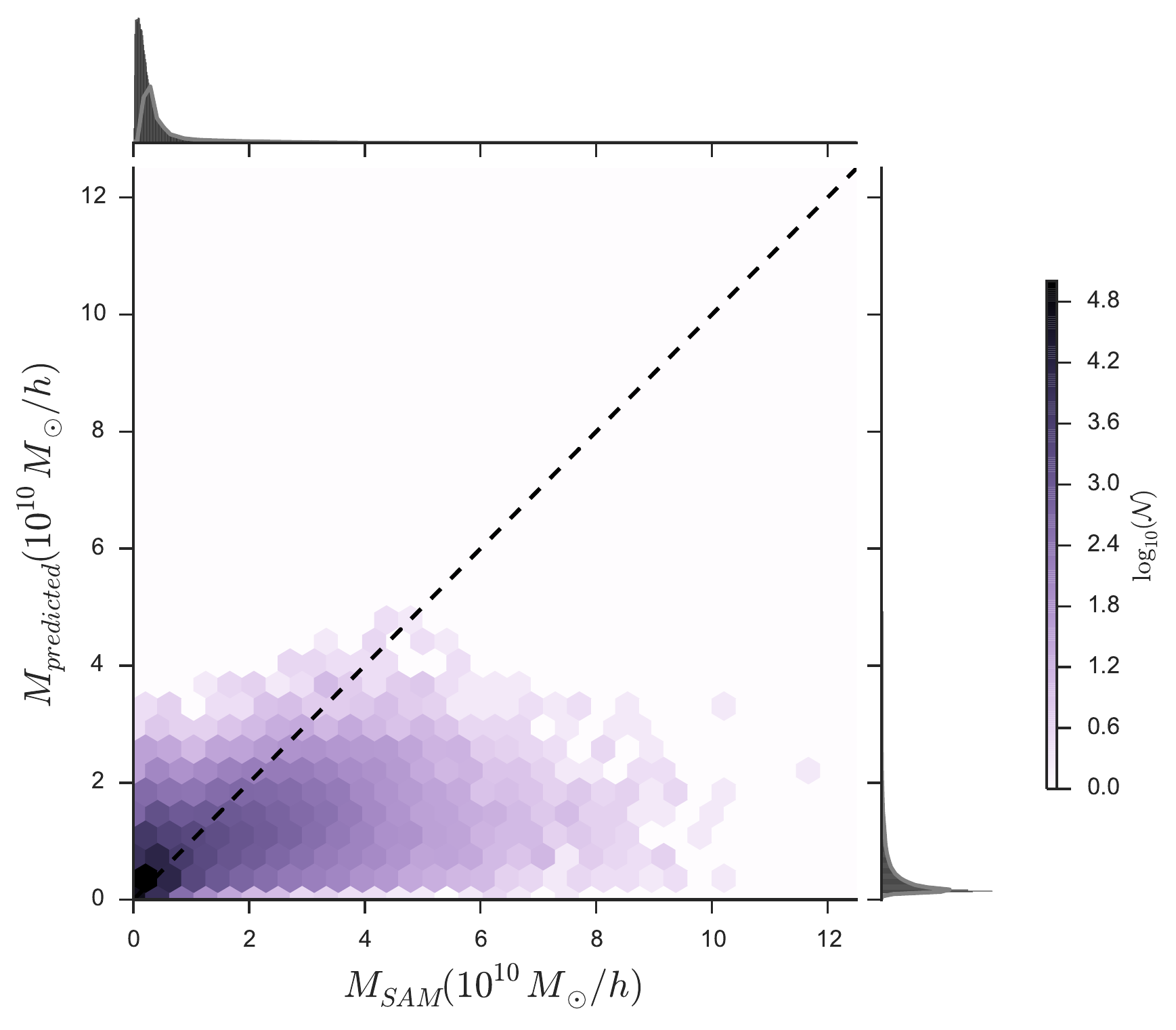}
\caption{A hexbin plot of $M_{SAM,cold}$ and $M_{predicted,cold}$. The black dashed line corresponds to a perfect prediction. The MSE for the prediction is 0.319, the Pearson correlation between the predicted galaxy mass and the G11 galaxy mass is 0.632 and the regression score is 0.40.}
\label{cold1}

\end{figure}

\begin{figure}
  \includegraphics[width=84mm]{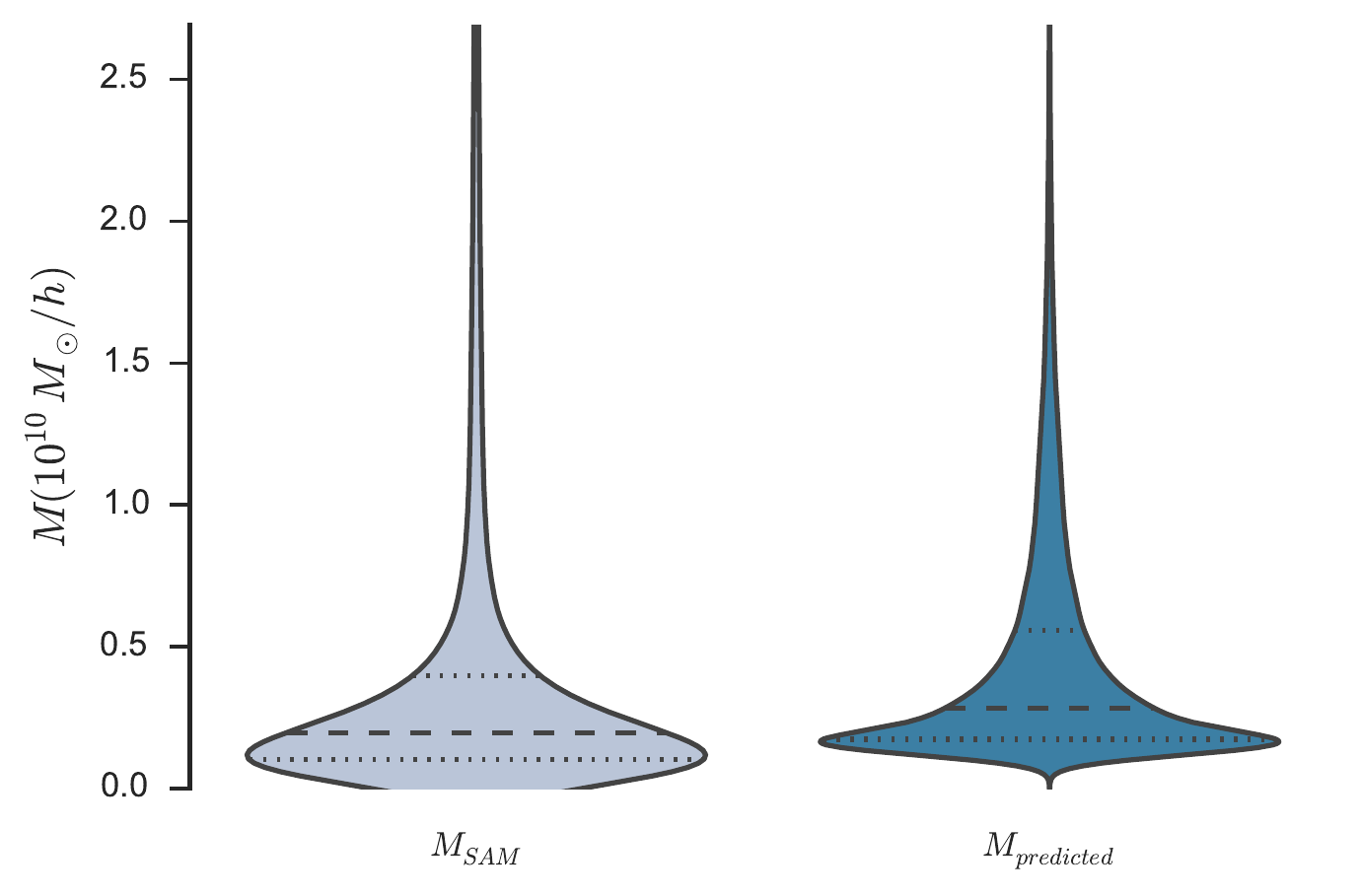}
  \caption{A violinplot showing the distributions of  $M_{SAM,cold}$ and $M_{predicted,cold}$. The median and the interquartile range for both sets of galaxy masses are shown.}
    \label{cold2}

\end{figure}

\begin{figure}
\includegraphics[width=84mm]{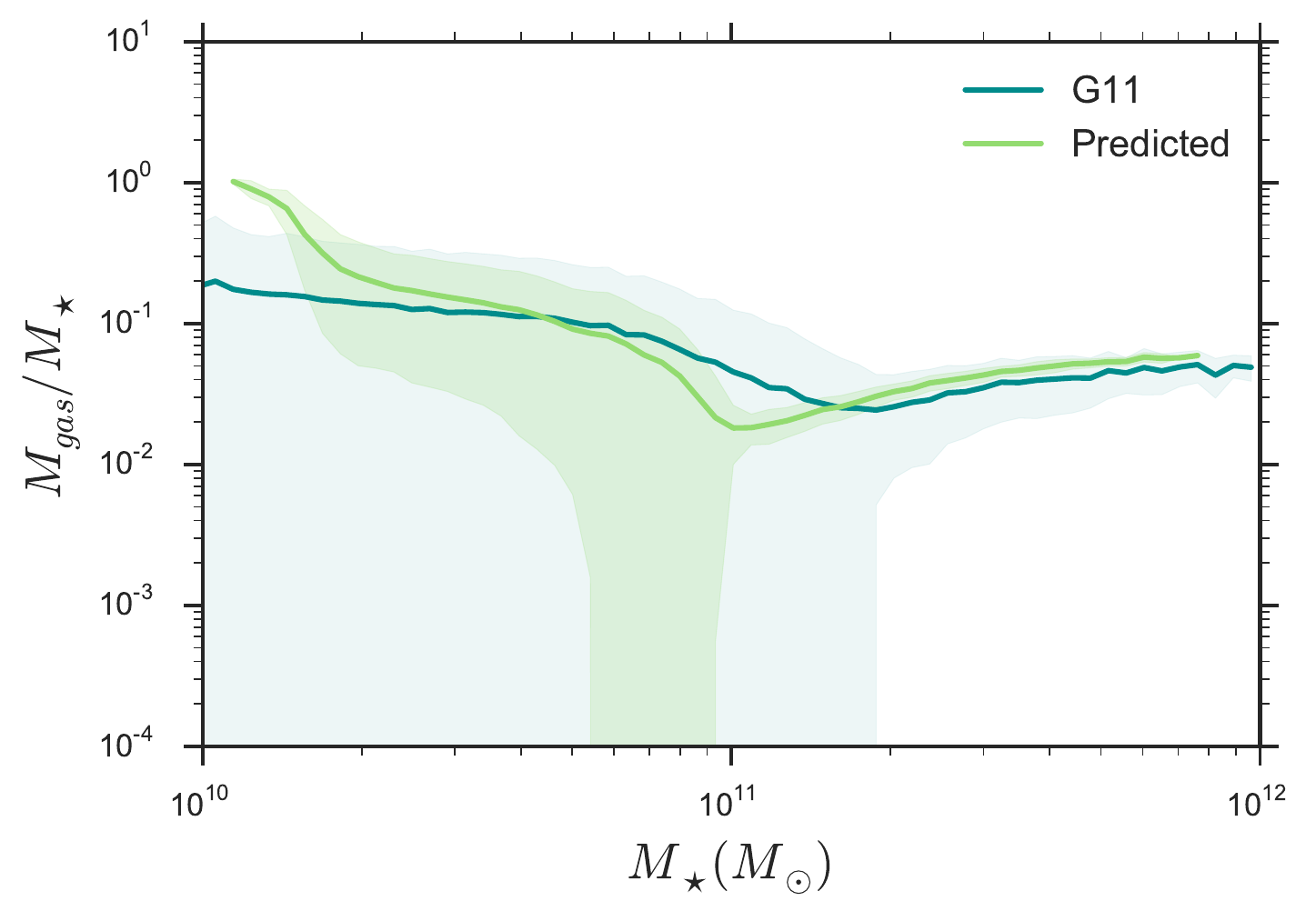}
\caption{The average cold gas mass fraction as a function of stellar mass is shown for G11 galaxies and ML galaxies. For G11, $\frac{M_{cold,SAM}}{M_{\star,SAM}}$ points are binned by $M_{\star,SAM}$ and for ML, $\frac{M_{cold,predicted}}{M_{\star,predicted}}$ are binned by $M_{\star,predicted}$. }
\label{gasfrac}

\end{figure}

\begin{figure}
  \includegraphics[width=84mm]{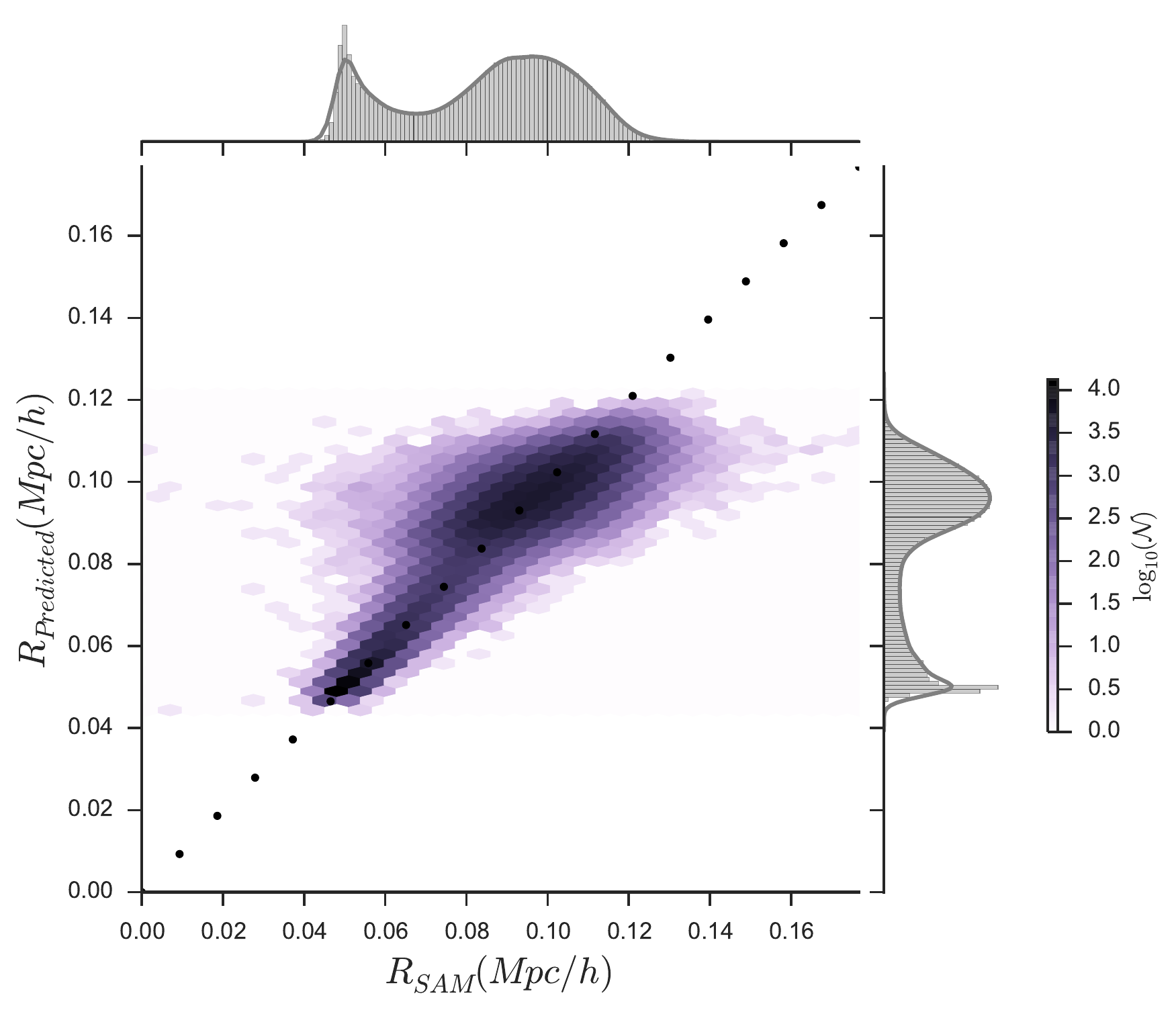}
  \caption{A hexbin plot of the cooling radii $R_{SAM,cool}$ and $R_{predicted,cool}$. The black dashed line corresponds to a perfect prediction. The MSE for the prediction is 0.000059, the Pearson cors between the predicted galaxy mass and the G11 galaxy mass is 0.930 and the regression score is 0.87.}
    \label{radius1}

\end{figure}

\begin{figure}
  \includegraphics[width=84mm]{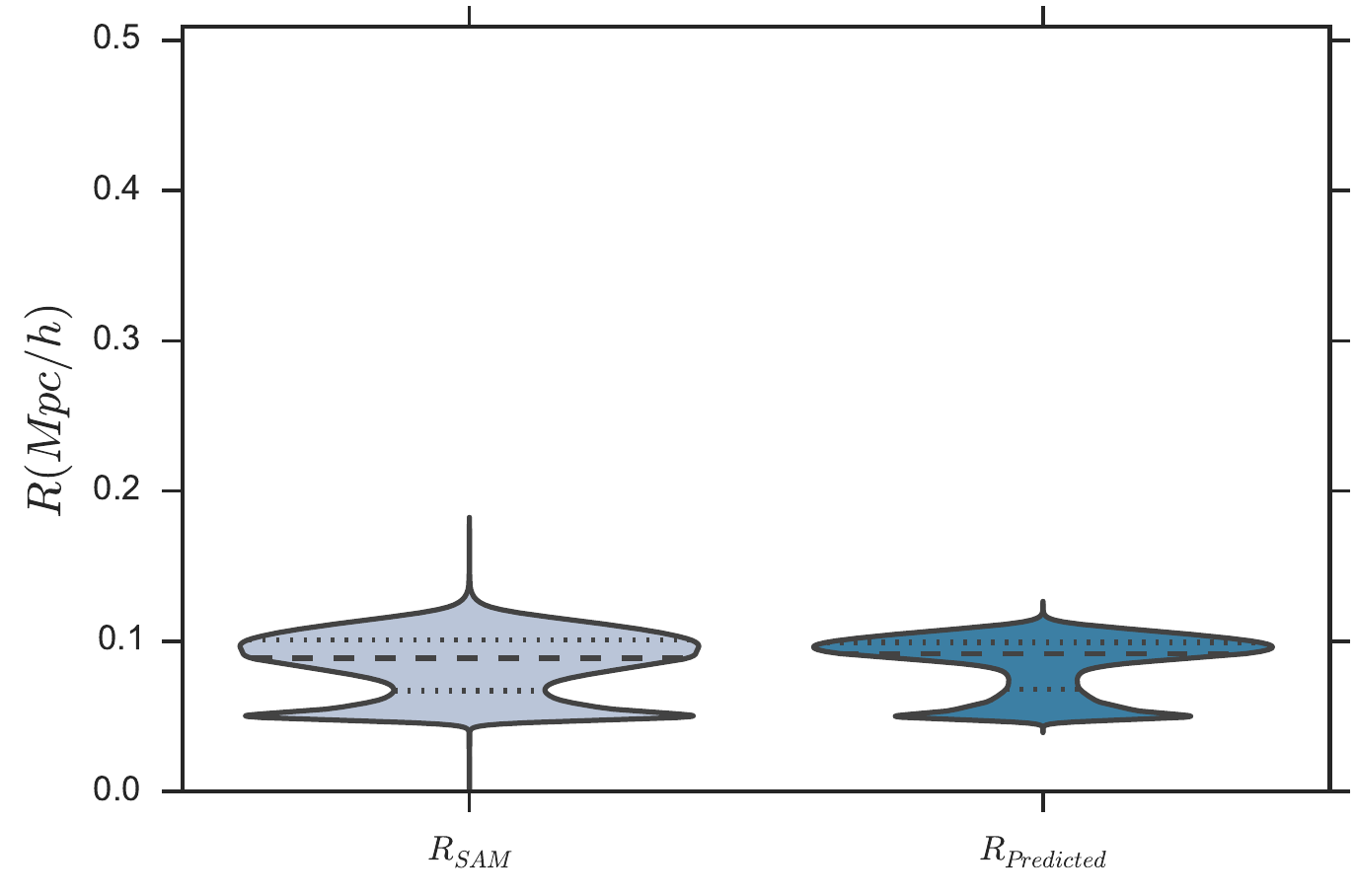}
  \caption{A violinplot showing the distributions of the $R_{SAM,cooling}$ and $R_{predicted,cooling}$. The median and the interquartile range for both sets of cooling radii are shown.}
    \label{radius2}

\end{figure}

\begin{figure}
  \includegraphics[width=84mm]{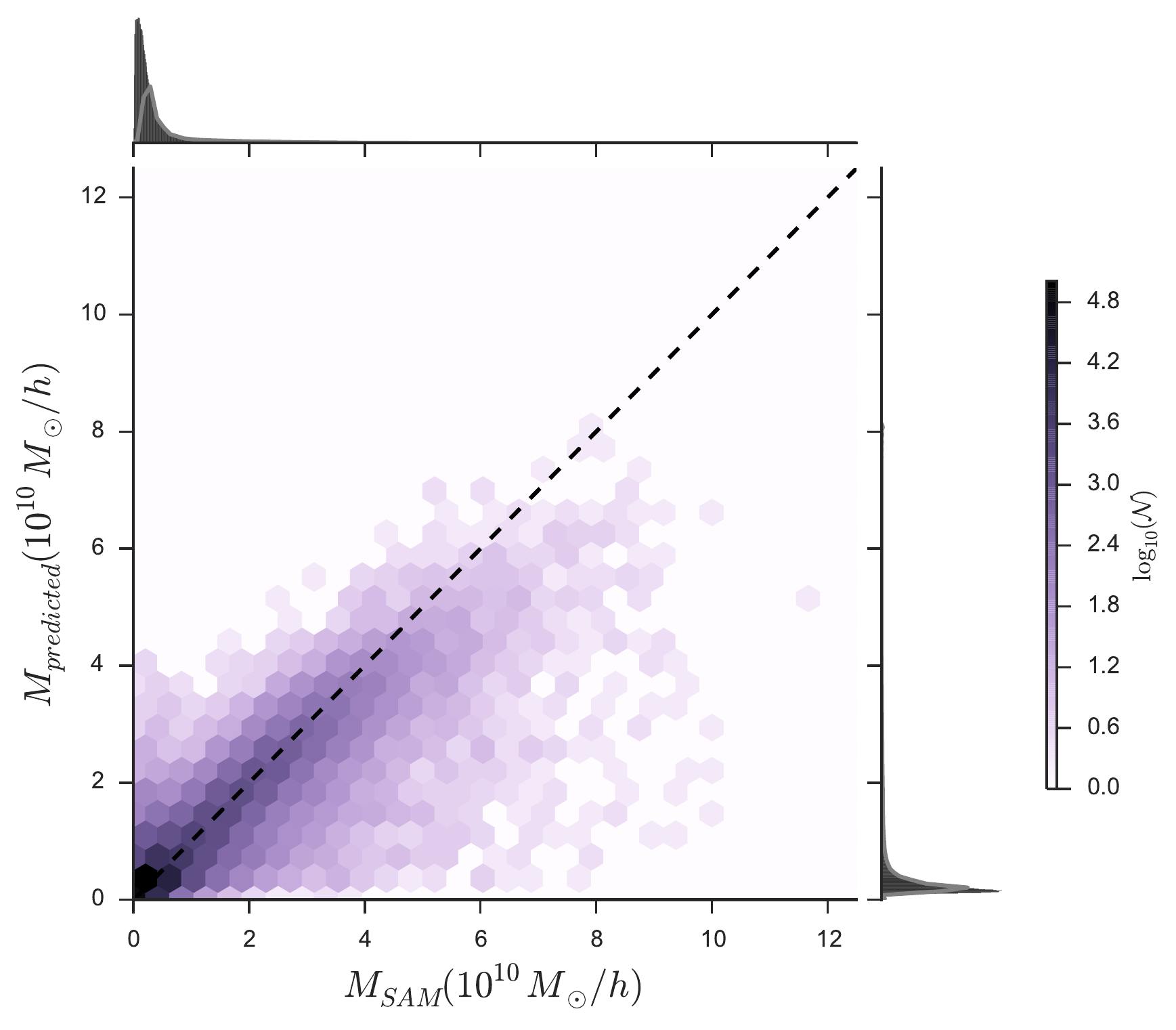}
  \caption{ A hexbin plot of $M_{SAM,cold}$ and $M_{predicted,cold}$. The black dashed line corresponds to a perfect prediction. The MSE for the prediction is 0.097, the Pearson correlation between the predicted galaxy mass and the G11 galaxy mass is 0.905 and the regression score is 0.82. The black dotted line corresponds to a perfect prediction. The main difference between this figure and figure 11 is that the cooling radius and the hot gas mass for two snapshots are explicitly included in the inputs for the ML algorithms.}
  \label{cold_b1}

\end{figure}

\begin{figure}
  \includegraphics[width=84mm]{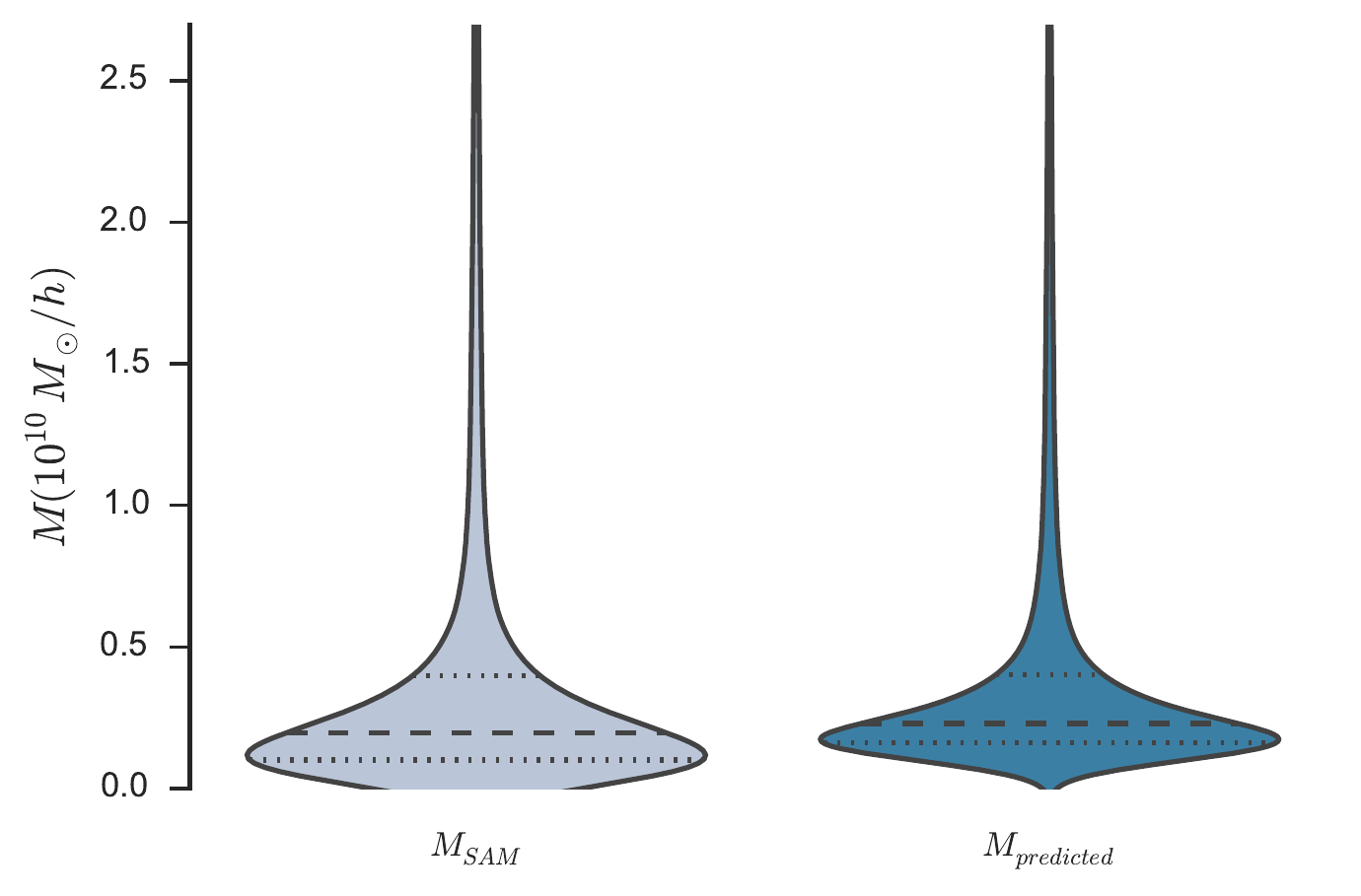}
  \caption{A violinplot showing the distributions of  $M_{SAM,cold}$ and $M_{predicted,cold}$. The median and the interquartile range for both sets of galaxy masses are shown. The main difference here is that the cooling radius and the hot gas mass for two snapshots are explicitly included in the inputs for the ML algorithms.}
    \label{cold_b2}

\end{figure}

\subsection{Cold Gas Mass} \label{results_cold}
As shown in Figure \ref{cold1}, the cold gas mass is being visibly underpredicted. This underprediction is unfortunately not surprising. The recipe used in G11, outlined in Section \ref{mdf: db_sam_cold}, has a partial halo dependence but the baryonic processes play a far more important role in determining the cold gas mass. Indeed, NW found no easy way to parameterize the efficiency of cooling rate in terms of the host halo mass and time and had to empirically estimate this value by running DLB07 on the mini-Millennium simulation. To investigate possible reasons for the underprediction, recall that the cooling rate is heavily dependent upon the cooling radius. By using the same algorithms and the same inputs, we predicted the cooling radius at $z=0$ for the central galaxy. Our results are shown in Figures \ref{radius1} and \ref{radius2}. We obtained a regression score of $0.86$ and a correlation of 0.931; thus, the prediction is fairly robust. The cooling radius, as shown in Equation \ref{rcool}, also has only a partial halo dependence. It is remarkable that our prediction is so accurate, then, using solely dark matter halo information; one would expect that baryonic recipes, for instance, the cooling function prescribed in \citet{sutherland1993cooling}, to play a far larger role in the determination of the cooling radius.

\begin{table*}
\begin{minipage}{159mm}
\caption{Cold gas mass prediction with the inclusion of cooling radius and hot gas mass as inputs}
 \label{cold_mse}
 \begin{tabular}{@{}lcccccc}
  \hline
  \hline
   & Technique & $MSE_b$ & $MSE$ & Factor ($\frac{MSE_b}{MSE}$) & Pearson Correlation
         & $R^2$\\
  \hline
  & Random Forests &  & \textcolor{darkgreen}{\textbf{\textit{0.319}}} & 1.652 & 0.631 & \textcolor{darkgreen}{\textbf{\textit{0.395}}} \\
  \multirow{-2}{*}{$M_{cold,w/o}$} & \textcolor{darkgreen}{\textbf{Extremely Randomized Trees}} & \multirow{-2}{*}{0.527} & \textcolor{darkgreen}{\textit{\textbf{0.319}}} & \textcolor{darkgreen}{\textbf{\textit{1.654}}} & \textcolor{darkgreen}{\textbf{\textit{0.632}}} & \textcolor{darkgreen}{\textbf{\textit{0.395}}}\\
  \hline 
  & \textcolor{darkgreen}{\textbf{Random Forests}} &  & \textcolor{darkgreen}{\textbf{\textit{0.0972}}} & \textcolor{darkgreen}{\textbf{\textit{5.422}}} & \textcolor{darkgreen}{\textbf{\textit{0.905}}} & \textcolor{darkgreen}{\textbf{\textit{0.816}}}\\
  \multirow{-2}{*}{$M_{cold,with}$} & Extremely Randomized Trees & \multirow{-2}{*}{0.527}& 0.113 & 4.664 & 0.892 & 0.786 \\
  \hline
 \end{tabular}
 \end{minipage}
\end{table*}

\par The reproduction of the cooling radius and the robust hot gas mass prediction discussed earlier raise the question: why is the cold gas mass evolution not being captured? ML is able to predict the two basic ingredients of gas cooling well, but the cooled mass itself is not being robustly predicted. We hypothesize that this discrepancy is a result of the variability in the cooling radius. Without some form of explicit inclusion of the time evolution of hot gas mass and cooling radius over snapshots, ML is unable to capture the evolution of the cooling radius and, consequently, the accumulated cold gas mass. We tested this hypothesis by including the cooling radius and the hot gas mass over only two snapshots ($z=0$ and $z=0.012$) in our inputs for ERT and RTF and repeating the cold gas mass prediction with the same parameters. 

\par Our results are presented in Figures \ref{cold_b1} and \ref{cold_b2} and Table \ref{cold_mse}. By just including these four additional inputs, our predictions became significantly more robust. Our regression score more than doubled and we can see in Figures \ref{cold_b2} that the distribution from G11 is being fairly reproduced. The comparison of the two distributions (figure 13 and 18) shows that the prescription used for evolving the cooling ODE is being fairly well picked up by the additional baryonic inputs included. This result is particularly striking because the cooling ODE was evolved 20 times between each snapshot in the Munich SAMs \citep{de2010semi, knebe2015nifty} and we included results only at the two extreme points. The improved results imply that ML is strong at extrapolating the underlying physical process in G11 for cooling only if baryonic ingredients are partially, explicitly included in the inputs to guide the algorithms. Furthermore, our results also instill confidence in our hypothesis that ML is unable to predict the accumulated cold gas mass because of its inability to pick up on the cooling evolution using solely dark matter inputs. 

\par 
Moreover, from both Tables \ref{mse} and Figures \ref{cold1} and \ref{cold2}, we can see that the machine learning algorithms are definitely learning partial information about how cold gas is being accumulated in G11. The distribution, while not a perfect fit by any means, does reproduce a narrower peak at about the same mass and the median and interquartile range are appreciably similar. Moreover Table \ref{mse} shows that the Pearson correlation between the predicted and the test set is 0.63 and $R^2$ is 0.40. We also see in the hexbin plot that the densest hexbins lie on the straight line, but there is  noticeable scatter for higher cold gas masses. Furthermore, in Figure \ref{gasfrac}, we show the average cold gas fraction as a function of stellar mass for both G11 galaxies ML galaxies. There are some discrepancies between our results and G11's at the beginning, but the general progression of the two curves matches up quite well. Our relatively poor results for the cold gas mass confirm previous results in literature by demonstrably and quantitatively showing the absence of a relatively simple mapping between cold gas mass and dark matter halo properties. 

\par 
While the main point of this study was to explore the halo-galaxy connection in a dark matter-only context, we are able to show the power of ML in modeling the complicated cold gas mass recipe reasonably well when very crude, partial baryonic ingredients are included as inputs. The analysis of cold gas mass accumulation using both dark matter only inputs and with baryonic inputs raises two interesting points; first, ML, by itself, is unable to pick up on the baryonic evolution involved in gas cooling using solely dark matter inputs. Second, with just the addition of four baryonic inputs, the prediction for the cold gas mass is vastly improved. The results above show that there is more room for exploration and that the relatively poor cold gas mass prediction does not undermine the overall usefulness of ML as a solid tool in exploring the problem of galaxy formation. 
 
\section{General Discussion} \label{discussion}
\par
The results above show that ML is able to recreate a population of galaxies that is strikingly similar to that of G11's in our dark matter-only framework. While the reduced MSE's we found for the different components of the mass are surprisingly low, they're still high enough to merit a discussion of the sources of error. First, and most important, a reason for the relatively high MSE is the absence of any baryonic processes or results being input into our machine learning algorithms. In NW, only the efficiency of the physical processes were modeled by using the host halo mass and redshift; NW did include simplified, but still physically motivated, baryonic processes in the model, with hand-tuned free parameters. Our model, on the other hand, is not grounded on physical motivations related to baryons at all; it is instead an effort to explore the halo-galaxy connection in the framework of SAMs by using solely halo properties and a partial merger history. Another possible reason for the relatively high MSE may be a result of the fact that we are only looking at the mass of the central galaxy of a halo and ignoring the satellite galaxies, while using the inputs for the entire halo. This may also explain some of the scatter we see in our results for the stellar mass and the cold gas mass. 

\par 
There are certain deficiencies in our model. As mentioned earlier, the model is not motivated by baryonic physics like most SAMs. G11 and other SAMs offer simple, but incredibly illuminating treatments of interactions at the galactic scale. Machine learning, on the other, by its very nature, does not leave room for exploration into the subtleties of baryonic physics and is a purely phenomenological model. Moreover, the predictions shown above do not imply that machine learning is a replacement to SAMs. Our results imply instead that machine learning offers an interesting and promising avenue for exploration in the domain of galaxy formation, primarily because of its simplicity, efficiency and its ability to provide a unique platform that allows us to probe how much information can be extracted from just dark matter haloes. In the case of the stellar mass, black hole mass and hot gas mass, ML is able to pick up on the physical prescriptions used by G11 very well by using solely a partial merger history and information about the halo environment. In the case of a cold gas mass, ML is not able to pick up on the evolution of gas cooling by itself; with a partial inclusion of a couple of baryonic inputs over two snapshots, however, we're able to double our regression score and make predictions that are vastly more robust. The improved results place confidence in the predictive power of ML and imply that ML may be a useful tool for future studies of NBHS and other problems in theoretical astrophysics. 

\par An interesting point here is the superficial similarity between our model and suhalo abundance matching (SHAM) \citep{conroy2009connecting}. Both models use halo information to glean physical information about the galaxies residing in the halo. Our study differs from SHAMs in one very key aspect: SHAM involves populating haloes with galaxies assuming that there already exists a monotonic relationship between halo mass and galaxy stellar mass (or luminosity). On the other hand, our model predicts properties of galaxies that have already been populated using a SAM (G11) with no relationship being fed to the algorithms. The results obtained, then, imply that the key assumption in most SHAMs that observable properties of galaxies are monotonically related to the dynamical properties of dark matter substructures is partially valid. The discrepancy in our cold gas mass result implies that the baryonic physics plays a vastly more important role in the cooling rate than the halo environment itself. But the reproduction of the total stellar mass implies that SHAM's assumptions in the context of stellar mass hold true and instill further confidence into the general methodology of SHAM. 

\par 
The cold gas mass prediction raises several interesting points. First, and most important, similar to \citet{contreras2015galaxy}, only a weak mapping between the cold gas mass and the internal halo properties was found. However, the robust prediction of the cooling radius and the hot gas mass does leave the door open for further exploration into modeling cold gas mass using ML. Our results also quantitatively verify what NW found; they were unable to parameterize the cooling efficiency in terms of host halo mass and the redshift. By the inclusion of just the cooling radius and hot gas mass over two snapshots, ML is able to vastly improve upon the DM-only predictions. The improved predictions naturally raise the question: can ML be applied to other, more complicated evolutionary models and still reproduce physically and numerically reasonable results? 

\par 
Our results give a unique and deeper look into the galaxy-halo connection that is grounded upon SAMs. We are able to quantitatively estimate the amount of information that dark matter haloes and merger trees hold about the baryonic processes that drive galaxy formation and evolution, in the context of SAMs and how SAMs populate haloes with galaxies. We have quantitatively shown that the environmental dependence of galaxy evolution on the surrounding dark matter halo is surprisingly strong. As mentioned earlier, our robust predictions for the total stellar mass, stellar mass in the bulge, black hole mass, and the hot gas mass strongly imply that it is possible to learn the physical processes used in cutting edge SAMs to evolve these components by using solely dark matter properties, a merger history, and machine learning. The relatively weaker results for the cold gas mass imply that our phenomonological, dark matter-only model fails in reproducing the cold gas mass evolution in galaxies. However, our improved cooling model with baryonic inputs solidifies the viability of ML in future galaxy formation studies. 

\par
Overall, we get somewhat surprising results since one would expect that gaseous interactions play a significantly more important role in predicting the final components of mass of a single galaxy than just the basic dark matter halo model; but, we have shown that machine learning provides a unique and fairly robust avenue to quantitatively analyze the role that just the dark matter haloes play in galaxy formation in the context of SAMs. We showed that the SMHM relation is reproduced almost perfectly, the shapes of the predicted and true distributions of the different mass components are very similar, the BH mass-bulge mass relation is reproduced, and the cold gas mass fraction as a function of stellar mass is reasonably reproduced. Machine learning is able to learn an appreciable portion of the physical prescriptions used in G11 for galaxy formation using solely dark matter inputs. Moreover, the amount of time it took to run the whole pipeline took about three hours, considerably less than the hundreds or thousands of hours a typical SAM would require.

\section{Conclusions} \label{conclusion}
We have performed an extensive study of the halo-galaxy connection by using novel machine learning techniques in the backdrop of a state of the art SAM. Using G11 to train our ML algorithms, the total stellar mass, stellar mass in the bulge, cold gas mass and hot gas mass in the Millennium simulation are predicted. ML provides a powerful framework to explore the problem of galaxy formation in part due to its relative simplicity, computational efficiency and its ability to model complex physical relationships. The discrepancies in and weaknesses of our phenomenological model were discussed and the reasons for some of our relatively less robust predictions were also discussed. An improved cooling model with four additional baryonic inputs was implemented, which made the cold gas mass predictions significantly better and solidified ML's position as a model that can be used to probe the halo-galaxy connection in more detail, perhaps with sophisticated NBHS. 
\par
Our primary conclusions are as follows:

\begin{enumerate}
 \renewcommand{\theenumi}{(\arabic{enumi})}
  \item Exploring the extent of the influence of dark matter haloes and its past environment on galaxy formation and evolution is a non-trivial problem with poorly defined inputs and mappings. Semi-analytic modeling is the prevalent galaxy formation modeling technique that uses simple, yet physically powerful, recipes to populate dark matter haloes with galaxies. However, there is no clear way to explore the extent of the influence of dark matter haloes on the halo-galaxy solely by using just SAMs. Machine learning, on the other hand, provides an interesting alternative to standard techniques for three main reasons: powerful predictive capabilities, simplicity and efficiency. 
  
 \item By using the Millennium simulation and G11, we set up a model that used internal halo properties ($\mathcal{N}$, spin, $M_{crit200}$, $v_{max}$ and $\sigma_v$) and a partial merger history to predict different mass components of the central galaxy in each dark matter halo at $z=0$. No baryonic processes were incorporated in our initial analysis. We applied several sophisticated algorithms (kNN, regression trees, random forests and extremely randomized trees) to the Millennium data and we were able to reproduce a similar galaxy population. 
 
 \item The total stellar mass and the stellar mass in the bulge are predicted very well. The predicted and true distributions for both are almost identical. ML is able to model the physical prescriptions laid out in G11 for galactic stellar mass evolution. The stellar mass-halo mass relation that G11 found is recreated almost perfectly with some very minor discrepancies for $M_h \approx 10^{15} M_{\odot} h^{-1}$. The bulge mass prediction is also fairly robust, with the distributions being remarkably consistent. However, the bulge mass is slightly overpredicted for lower masses. We hypothesize that this may be because our inputs include a partial merger history of the haloes and not galaxies. Consequently, ML is possibly overpredicting as a result of its inability to fully model the galaxy-galaxy merger timescale using only a halo merger history. The central black hole mass prediction is also very robust and the distribution is recreated almost perfectly. The BH-bulge mass relation for the ML simulated galaxies and G11 galaxies is very consistent. There is a slight overprediction for lower masses for the central black hole mass prediction, which further places confidence in the hypothesis that ML is unable to fully pick up on the galaxy-galaxy merger timescale using only a halo merger history. 
 
 \item The hot gas mass is predicted outstandingly well. ML is demonstrably able to model G11's prescriptions for gas stripping and supernovae feedback. The cold gas mass prediction, on the other hand, is relatively weak with a correlation of only $0.63$. However, the robust cooling radius and the hot gas prediction imply that the ingredients for cooling are sufficiently modeled by ML. We hypothesized that our poor prediction was a result of the inability of ML to model the cooling radius evolution without any baryonic guidance (i.e. by using solely dark matter inputs and merger history). We tested this hypothesis by including the cooling radius and the hot gas mass for only the last two snapshots and found significantly better predictions with a correlation of $0.91$ and $R^2$ of $0.82$. The improved cooling predictions place confidence in the predictive power of ML and imply that ML will be a useful tool in future studies in galaxy formation and evolution. The average cold gas mass fraction as a function of stellar mass was also plotted for G11 galaxies and the predicted galaxies. The shape of the two curves is reasonably similar with a minor discrepancy at the lowest masses. 
 
 \item Our results provide a unique framework to explore the galaxy-halo connection that is built by using SAMs. We are able to quantitatively estimate the amount of information that dark matter haloes and merger trees hold about the baryonic processes that drive galaxy formation and evolution, in the context of G11. Our robust predictions for the total stellar mass, stellar mass in the bulge, central black hole mass, and the hot gas mass strongly imply that it is possible to successfully model the physical prescriptions used in SAMs to evolve these mass components using solely dark matter properties, a merger history and machine learning. However, ML is unable to find a robust approximate mapping between the internal dark matter halo properties and the cold gas mass, like \citet{neistein2010degeneracy, faucher2011baryonic, contreras2015galaxy}. 
 
 \item ML is a phenomenological model and not a physical one, and, consequently, is not a replacement for SAMs. However, ML offers a solid and intriguing framework to explore the halo-galaxy connection with solid results comparable to G11, which conventional modeling techniques don't provide. 
\end{enumerate}

The results presented in this paper show the usefulness of ML in providing a solid framework to probe the halo-galaxy connection in the backdrop of SAMs. Future work includes exploring more sophisticated ML techniques to probe galaxy formation and evolution in NBHS. 
\section*{Acknowledgments}

The authors thank Christopher Chan, Rishabh Jain, and Dingcheng Yue for help in gathering data and exploring preliminary machine learning approaches. HMK and RJB acknowledge support from the National Science Foundation Grant No. AST-1313415. HMK has been supported in part by funding from the LAS Honors Council at the University of Illinois and by the Office of Student Financial Aid at the University of Illinois. RJB has been supported in part by the Center for Advanced Studies at the University of Illinois. MJT is supported by the Gordon and Betty Moore Foundation's Data-Driven Discovery Initiative through Grant GBMF4561. We would like to thank the reviewer for their helpful comments that made this paper better. 

The Millennium Simulation databases used in this paper and the web application providing online access to them were constructed as part of the activities of the German Astrophysical Virtual Observatory (GAVO).

\footnotesize{
	\bibliographystyle{mnras}
	\bibliography{main}

\begin{thebibliography}{}
\makeatletter
\relax
\def\mn@urlcharsother{\let\do\@makeother \do\$\do\&\do\#\do\^\do\_\do\%\do\~}
\def\mn@doi{\begingroup\mn@urlcharsother \@ifnextchar [ {\mn@doi@}
  {\mn@doi@[]}}
\def\mn@doi@[#1]#2{\def\@tempa{#1}\ifx\@tempa\@empty \href
  {http://dx.doi.org/#2} {doi:#2}\else \href {http://dx.doi.org/#2} {#1}\fi
  \endgroup}
\def\mn@eprint#1#2{\mn@eprint@#1:#2::\@nil}
\def\mn@eprint@arXiv#1{\href {http://arxiv.org/abs/#1} {{\tt arXiv:#1}}}
\def\mn@eprint@dblp#1{\href {http://dblp.uni-trier.de/rec/bibtex/#1.xml}
  {dblp:#1}}
\def\mn@eprint@#1:#2:#3:#4\@nil{\def\@tempa {#1}\def\@tempb {#2}\def\@tempc
  {#3}\ifx \@tempc \@empty \let \@tempc \@tempb \let \@tempb \@tempa \fi \ifx
  \@tempb \@empty \def\@tempb {arXiv}\fi \@ifundefined
  {mn@eprint@\@tempb}{\@tempb:\@tempc}{\expandafter \expandafter \csname
  mn@eprint@\@tempb\endcsname \expandafter{\@tempc}}}

\bibitem[\protect\citeauthoryear{Angulo, Springel, White, Jenkins, Baugh  \&
  Frenk}{Angulo et~al.}{2012}]{angulo2012scaling}
Angulo R.,  Springel V.,  White S.,  Jenkins A.,  Baugh C.,   Frenk C.,  2012,
  MNRAS, 426, 2046

\bibitem[\protect\citeauthoryear{Baldry, Balogh, Bower, Glazebrook, Nichol,
  Bamford  \& Budavari}{Baldry et~al.}{2006}]{baldry2006galaxy}
Baldry I.~K.,  Balogh M.~L.,  Bower R.,  Glazebrook K.,  Nichol R.~C.,  Bamford
  S.~P.,   Budavari T.,  2006, MNRAS, 373, 469

\bibitem[\protect\citeauthoryear{Ball \& Brunner}{Ball \&
  Brunner}{2010}]{ball2010data}
Ball N.~M.,  Brunner R.~J.,  2010, International Journal of Modern Physics D,
  19, 1049

\bibitem[\protect\citeauthoryear{Ball, Brunner, Myers  \& Tcheng}{Ball
  et~al.}{2006}]{ball2006robust}
Ball N.~M.,  Brunner R.~J.,  Myers A.~D.,   Tcheng D.,  2006, ApJ, 650, 497

\bibitem[\protect\citeauthoryear{Ball, Brunner, Myers, Strand, Alberts, Tcheng
  \& Llor{\`a}}{Ball et~al.}{2007}]{ball2007robust}
Ball N.~M.,  Brunner R.~J.,  Myers A.~D.,  Strand N.~E.,  Alberts S.~L.,
  Tcheng D.,   Llor{\`a} X.,  2007, ApJ, 663, 774

\bibitem[\protect\citeauthoryear{Banerji et~al.,}{Banerji
  et~al.}{2010}]{banerji2010galaxy}
Banerji M.,  et~al., 2010, MNRAS, 406, 342

\bibitem[\protect\citeauthoryear{Baugh}{Baugh}{2006}]{baugh2006primer}
Baugh C.~M.,  2006, Reports on Progress in Physics, 69, 3101

\bibitem[\protect\citeauthoryear{Behroozi, Conroy  \& Wechsler}{Behroozi
  et~al.}{2010}]{behroozi2010comprehensive}
Behroozi P.~S.,  Conroy C.,   Wechsler R.~H.,  2010, ApJ, 717, 379

\bibitem[\protect\citeauthoryear{Benson}{Benson}{2012}]{benson2012galacticus}
Benson A.~J.,  2012, New Astronomy, 17, 175

\bibitem[\protect\citeauthoryear{Benson, Pearce, Frenk, Baugh  \&
  Jenkins}{Benson et~al.}{2001}]{benson2001comparison}
Benson A.,  Pearce F.,  Frenk C.,  Baugh C.,   Jenkins A.,  2001, MNRAS, 320,
  261

\bibitem[\protect\citeauthoryear{Blumenthal, Faber, Primack  \&
  Rees}{Blumenthal et~al.}{1984}]{blumenthal1984formation}
Blumenthal G.~R.,  Faber S.,  Primack J.~R.,   Rees M.~J.,  1984, Nature, 311,
  517

\bibitem[\protect\citeauthoryear{Bower, Benson, Malbon, Helly, Frenk, Baugh,
  Cole  \& Lacey}{Bower et~al.}{2006}]{bower2006breaking}
Bower R.,  Benson A.,  Malbon R.,  Helly J.,  Frenk C.,  Baugh C.,  Cole S.,
  Lacey C.~G.,  2006, MNRAS, 370, 645

\bibitem[\protect\citeauthoryear{Bower, Vernon, Goldstein, Benson, Lacey,
  Baugh, Cole  \& Frenk}{Bower et~al.}{2010}]{bower2010parameter}
Bower R.,  Vernon I.,  Goldstein M.,  Benson A.,  Lacey C.~G.,  Baugh C.,  Cole
  S.,   Frenk C.,  2010, MNRAS, 407, 2017

\bibitem[\protect\citeauthoryear{Boylan-Kolchin, Ma  \&
  Quataert}{Boylan-Kolchin et~al.}{2008}]{boylan2008dynamical}
Boylan-Kolchin M.,  Ma C.-P.,   Quataert E.,  2008, MNRAS, 383, 93

\bibitem[\protect\citeauthoryear{Breiman}{Breiman}{1996}]{breiman1996bagging}
Breiman L.,  1996, Machine learning, 24, 123

\bibitem[\protect\citeauthoryear{Breiman}{Breiman}{2001}]{breiman2001random}
Breiman L.,  2001, Machine learning, 45, 5

\bibitem[\protect\citeauthoryear{Breiman, Friedman, Stone  \& Olshen}{Breiman
  et~al.}{1984}]{breiman1984classification}
Breiman L.,  Friedman J.,  Stone C.~J.,   Olshen R.~A.,  1984, Classification
  and regression trees.
CRC press

\bibitem[\protect\citeauthoryear{Chabrier}{Chabrier}{2003}]{chabrier2003galactic}
Chabrier G.,  2003, Publications of the Astronomical Society of the Pacific,
  115, 763

\bibitem[\protect\citeauthoryear{Cole, Aragon-Salamanca, Frenk, Navarro  \&
  Zepf}{Cole et~al.}{1994}]{cole1994recipe}
Cole S.,  Aragon-Salamanca A.,  Frenk C.~S.,  Navarro J.~F.,   Zepf S.~E.,
  1994, MNRAS, 271, 781

\bibitem[\protect\citeauthoryear{Cole, Lacey, Baugh  \& Frenk}{Cole
  et~al.}{2000}]{cole2000hierarchical}
Cole S.,  Lacey C.~G.,  Baugh C.~M.,   Frenk C.~S.,  2000, MNRAS, 319, 168

\bibitem[\protect\citeauthoryear{Collaboration et~al.}{Collaboration
  et~al.}{2015}]{planck2015planck}
Collaboration P.,  et~al., 2015, arXiv preprint arXiv:1502.01589

\bibitem[\protect\citeauthoryear{Conroy \& Wechsler}{Conroy \&
  Wechsler}{2009}]{conroy2009connecting}
Conroy C.,  Wechsler R.~H.,  2009, ApJ, 696, 620

\bibitem[\protect\citeauthoryear{Contreras, Baugh, Norberg  \&
  Padilla}{Contreras et~al.}{2015}]{contreras2015galaxy}
Contreras S.,  Baugh C.,  Norberg P.,   Padilla N.,  2015, arXiv preprint
  arXiv:1502.06614

\bibitem[\protect\citeauthoryear{Croton et~al.,}{Croton
  et~al.}{2006}]{croton2006many}
Croton D.~J.,  et~al., 2006, MNRAS, 365, 11

\bibitem[\protect\citeauthoryear{Cucciati et~al.,}{Cucciati
  et~al.}{2012}]{cucciati2012comparison}
Cucciati O.,  et~al., 2012, A\&A, 548, A108

\bibitem[\protect\citeauthoryear{Davis, Efstathiou, Frenk  \& White}{Davis
  et~al.}{1985}]{davis1985evolution}
Davis M.,  Efstathiou G.,  Frenk C.~S.,   White S.~D.,  1985, ApJ, 292, 371

\bibitem[\protect\citeauthoryear{De~La~Torre et~al.,}{De~La~Torre
  et~al.}{2011}]{de2011comparison}
De~La~Torre S.,  et~al., 2011, A\&A, 525, A125

\bibitem[\protect\citeauthoryear{De~Lucia \& Blaizot}{De~Lucia \&
  Blaizot}{2007}]{de2007hierarchical}
De~Lucia G.,  Blaizot J.,  2007, MNRAS, 375, 2

\bibitem[\protect\citeauthoryear{De~Lucia, Kauffmann  \& White}{De~Lucia
  et~al.}{2004}]{de2004chemical}
De~Lucia G.,  Kauffmann G.,   White S.~D.,  2004, MNRAS, 349, 1101

\bibitem[\protect\citeauthoryear{De~Lucia, Springel, White, Croton  \&
  Kauffmann}{De~Lucia et~al.}{2006}]{de2006formation}
De~Lucia G.,  Springel V.,  White S.~D.,  Croton D.,   Kauffmann G.,  2006,
  MNRAS, 366, 499

\bibitem[\protect\citeauthoryear{De~Lucia, Boylan-Kolchin, Benson, Fontanot  \&
  Monaco}{De~Lucia et~al.}{2010}]{de2010semi}
De~Lucia G.,  Boylan-Kolchin M.,  Benson A.~J.,  Fontanot F.,   Monaco P.,
  2010, MNRAS, 406, 1533

\bibitem[\protect\citeauthoryear{Dieleman, Willett  \& Dambre}{Dieleman
  et~al.}{2015}]{dieleman2015rotation}
Dieleman S.,  Willett K.~W.,   Dambre J.,  2015, MNRAS, 450, 1441

\bibitem[\protect\citeauthoryear{Faucher-Gigu{\`e}re, Kere{\v{s}}  \&
  Ma}{Faucher-Gigu{\`e}re et~al.}{2011}]{faucher2011baryonic}
Faucher-Gigu{\`e}re C.-A.,  Kere{\v{s}} D.,   Ma C.-P.,  2011, MNRAS, 417, 2982

\bibitem[\protect\citeauthoryear{Fiorentin, Bailer-Jones, Lee, Beers, Sivarani,
  Wilhelm, Prieto  \& Norris}{Fiorentin et~al.}{2007}]{fiorentin2007estimation}
Fiorentin P.~R.,  Bailer-Jones C.,  Lee Y.,  Beers T.,  Sivarani T.,  Wilhelm
  R.,  Prieto C.~A.,   Norris J.,  2007, A\&A, 467, 1373

\bibitem[\protect\citeauthoryear{Gerdes, Sypniewski, McKay, Hao, Weis, Wechsler
   \& Busha}{Gerdes et~al.}{2010}]{gerdes2010arborz}
Gerdes D.~W.,  Sypniewski A.~J.,  McKay T.~A.,  Hao J.,  Weis M.~R.,  Wechsler
  R.~H.,   Busha M.~T.,  2010, ApJ, 715, 823

\bibitem[\protect\citeauthoryear{Geurts, Ernst  \& Wehenkel}{Geurts
  et~al.}{2006}]{geurts2006extremely}
Geurts P.,  Ernst D.,   Wehenkel L.,  2006, Machine learning, 63, 3

\bibitem[\protect\citeauthoryear{Graff, Feroz, Hobson  \& Lasenby}{Graff
  et~al.}{2014}]{graff2014skynet}
Graff P.,  Feroz F.,  Hobson M.~P.,   Lasenby A.,  2014, MNRAS, 441, 1741

\bibitem[\protect\citeauthoryear{Guo et~al.,}{Guo et~al.}{2011}]{guo2011dwarf}
Guo Q.,  et~al., 2011, MNRAS, 413, 101

\bibitem[\protect\citeauthoryear{Henriques, Thomas, Oliver  \&
  Roseboom}{Henriques et~al.}{2009}]{henriques2009monte}
Henriques B.~M.,  Thomas P.~A.,  Oliver S.,   Roseboom I.,  2009, MNRAS, 396,
  535

\bibitem[\protect\citeauthoryear{Henriques, White, Thomas, Angulo, Guo, Lemson
  \& Springel}{Henriques et~al.}{2013}]{henriques2013simulations}
Henriques B.~M.,  White S.~D.,  Thomas P.~A.,  Angulo R.~E.,  Guo Q.,  Lemson
  G.,   Springel V.,  2013, MNRAS, 431, 3373

\bibitem[\protect\citeauthoryear{Hopkins et~al.,}{Hopkins
  et~al.}{2010}]{hopkins2010mergers}
Hopkins P.~F.,  et~al., 2010, ApJ, 715, 202

\bibitem[\protect\citeauthoryear{Ivezi{\'c}, Connolly, VanderPlas  \&
  Gray}{Ivezi{\'c} et~al.}{2014}]{ivezic2014statistics}
Ivezi{\'c} {\v{Z}}.,  Connolly A.~J.,  VanderPlas J.~T.,   Gray A.,  2014,
  Statistics, Data Mining, and Machine Learning in Astronomy: A Practical
  Python Guide for the Analysis of Survey Data: A Practical Python Guide for
  the Analysis of Survey Data.
Princeton University Press

\bibitem[\protect\citeauthoryear{Johnson \& Zhang}{Johnson \&
  Zhang}{2011}]{johnson2011learning}
Johnson R.,  Zhang T.,  2011, arXiv preprint arXiv:1109.0887

\bibitem[\protect\citeauthoryear{Kamdar, Turk  \& Brunner}{Kamdar
  et~al.}{tted}]{kamdar2015machine}
Kamdar H.,  Turk M.,   Brunner R.,  Submitted, MNRAS

\bibitem[\protect\citeauthoryear{Kang, Jing, Mo  \& B{\"o}rner}{Kang
  et~al.}{2005}]{kang2005semianalytical}
Kang X.,  Jing Y.,  Mo H.,   B{\"o}rner G.,  2005, ApJ, 631, 21

\bibitem[\protect\citeauthoryear{Kauffmann}{Kauffmann}{1996}]{kauffmann1996disc}
Kauffmann G.,  1996, MNRAS, 281, 475

\bibitem[\protect\citeauthoryear{Kauffmann, White  \& Guiderdoni}{Kauffmann
  et~al.}{1993}]{kauffmann1993formation}
Kauffmann G.,  White S.~D.,   Guiderdoni B.,  1993, MNRAS, 264, 201

\bibitem[\protect\citeauthoryear{Kennicutt~Jr}{Kennicutt~Jr}{1998}]{kennicutt1998global}
Kennicutt~Jr R.~C.,  1998, ApJ, 498, 541

\bibitem[\protect\citeauthoryear{Kim}{Kim}{2015}]{kim2015hybrid}
Kim Edward B. R. C.-K.~M.,  2015, MNRAS, 453, 507

\bibitem[\protect\citeauthoryear{Kind \& Brunner}{Kind \&
  Brunner}{2013}]{kind2013tpz}
Kind M.~C.,  Brunner R.~J.,  2013, MNRAS, 432, 1483

\bibitem[\protect\citeauthoryear{Klypin, Trujillo-Gomez  \& Primack}{Klypin
  et~al.}{2011}]{klypin2011dark}
Klypin A.~A.,  Trujillo-Gomez S.,   Primack J.,  2011, ApJ, 740, 102

\bibitem[\protect\citeauthoryear{Knebe et~al.,}{Knebe
  et~al.}{2015}]{knebe2015nifty}
Knebe A.,  et~al., 2015, arXiv preprint arXiv:1505.04607

\bibitem[\protect\citeauthoryear{Lagos, Cora  \& Padilla}{Lagos
  et~al.}{2008}]{lagos2008effects}
Lagos C. d.~P.,  Cora S.~A.,   Padilla N.~D.,  2008, MNRAS, 388, 587

\bibitem[\protect\citeauthoryear{Lemson et~al.}{Lemson
  et~al.}{2006}]{lemson2006halo}
Lemson G.,  et~al., 2006, arXiv preprint astro-ph/0608019

\bibitem[\protect\citeauthoryear{Liu \& Weisberg}{Liu \&
  Weisberg}{2011}]{liu2011review}
Liu Y.,  Weisberg R.~H.,  2011, A review of self-organizing map applications in
  meteorology and oceanography.
INTECH Open Access Publisher

\bibitem[\protect\citeauthoryear{Liu, Yang, Mo, Van~den Bosch  \& Springel}{Liu
  et~al.}{2010}]{liu2010stellar}
Liu L.,  Yang X.,  Mo H.,  Van~den Bosch F.~C.,   Springel V.,  2010, ApJ, 712,
  734

\bibitem[\protect\citeauthoryear{Martin}{Martin}{1999}]{martin1999properties}
Martin C.~L.,  1999, ApJ, 513, 156

\bibitem[\protect\citeauthoryear{Mo, Mao  \& White}{Mo
  et~al.}{1998}]{mo1998formation}
Mo H.,  Mao S.,   White S.~D.,  1998, MNRAS, 295, 319

\bibitem[\protect\citeauthoryear{Monaco, Fontanot  \& Taffoni}{Monaco
  et~al.}{2007}]{monaco2007morgana}
Monaco P.,  Fontanot F.,   Taffoni G.,  2007, MNRAS, 375, 1189

\bibitem[\protect\citeauthoryear{Monaco, Benson, De~Lucia, Fontanot, Borgani
  \& Boylan-Kolchin}{Monaco et~al.}{2014}]{monaco2014semi}
Monaco P.,  Benson A.~J.,  De~Lucia G.,  Fontanot F.,  Borgani S.,
  Boylan-Kolchin M.,  2014, MNRAS, 441, 2058

\bibitem[\protect\citeauthoryear{Moster, Somerville, Maulbetsch, Van~den Bosch,
  Macci{\`o}, Naab  \& Oser}{Moster et~al.}{2010}]{moster2010constraints}
Moster B.~P.,  Somerville R.~S.,  Maulbetsch C.,  Van~den Bosch F.~C.,
  Macci{\`o} A.~V.,  Naab T.,   Oser L.,  2010, ApJ, 710, 903

\bibitem[\protect\citeauthoryear{Neistein \& Weinmann}{Neistein \&
  Weinmann}{2010}]{neistein2010degeneracy}
Neistein E.,  Weinmann S.~M.,  2010, MNRAS, 405, 2717

\bibitem[\protect\citeauthoryear{Ntampaka, Trac, Sutherland, Battaglia, Poczos
  \& Schneider}{Ntampaka et~al.}{2015}]{ntampaka2015machine}
Ntampaka M.,  Trac H.,  Sutherland D.~J.,  Battaglia N.,  Poczos B.,
  Schneider J.,  2015, ApJ, 803, 50

\bibitem[\protect\citeauthoryear{Pedregosa et~al.,}{Pedregosa
  et~al.}{2011}]{pedregosa2011scikit}
Pedregosa F.,  et~al., 2011, The Journal of Machine Learning Research, 12, 2825

\bibitem[\protect\citeauthoryear{Peebles}{Peebles}{1982}]{peebles1982large}
Peebles P.,  1982, Astrophys. J, 263, L1

\bibitem[\protect\citeauthoryear{Roe, Yang, Zhu, Liu, Stancu  \& McGregor}{Roe
  et~al.}{2005}]{roe2005boosted}
Roe B.~P.,  Yang H.-J.,  Zhu J.,  Liu Y.,  Stancu I.,   McGregor G.,  2005,
  Nuclear Instruments and Methods in Physics Research Section A: Accelerators,
  Spectrometers, Detectors and Associated Equipment, 543, 577

\bibitem[\protect\citeauthoryear{Schaye et~al.,}{Schaye
  et~al.}{2015}]{schaye2015eagle}
Schaye J.,  et~al., 2015, MNRAS, 446, 521

\bibitem[\protect\citeauthoryear{Silverman}{Silverman}{1986}]{silverman1986density}
Silverman B.~W.,  1986, Density estimation for statistics and data analysis.
 Vol. 26, CRC press

\bibitem[\protect\citeauthoryear{Skillman, Warren, Turk, Wechsler, Holz  \&
  Sutter}{Skillman et~al.}{2014}]{skillman2014dark}
Skillman S.~W.,  Warren M.~S.,  Turk M.~J.,  Wechsler R.~H.,  Holz D.~E.,
  Sutter P.,  2014, arXiv preprint arXiv:1407.2600

\bibitem[\protect\citeauthoryear{Somerville \& Dav{\'e}}{Somerville \&
  Dav{\'e}}{2014}]{somerville2014physical}
Somerville R.~S.,  Dav{\'e} R.,  2014, arXiv preprint arXiv:1412.2712

\bibitem[\protect\citeauthoryear{Somerville \& Primack}{Somerville \&
  Primack}{1999}]{somerville1999semi}
Somerville R.~S.,  Primack J.~R.,  1999, MNRAS, 310, 1087

\bibitem[\protect\citeauthoryear{Somerville, Hopkins, Cox, Robertson  \&
  Hernquist}{Somerville et~al.}{2008}]{somerville2008semi}
Somerville R.~S.,  Hopkins P.~F.,  Cox T.~J.,  Robertson B.~E.,   Hernquist L.,
   2008, MNRAS, 391, 481

\bibitem[\protect\citeauthoryear{Springel}{Springel}{2005}]{springel2005cosmological}
Springel V.,  2005, MNRAS, 364, 1105

\bibitem[\protect\citeauthoryear{Springel, White, Tormen  \&
  Kauffmann}{Springel et~al.}{2001}]{springel2001populating}
Springel V.,  White S.~D.,  Tormen G.,   Kauffmann G.,  2001, MNRAS, 328, 726

\bibitem[\protect\citeauthoryear{Springel et~al.,}{Springel
  et~al.}{2005}]{springel2005simulations}
Springel V.,  et~al., 2005, Nature, 435, 629

\bibitem[\protect\citeauthoryear{Sutherland \& Dopita}{Sutherland \&
  Dopita}{1993}]{sutherland1993cooling}
Sutherland R.~S.,  Dopita M.~A.,  1993, ApJ Supplement Series, 88, 253

\bibitem[\protect\citeauthoryear{Vogelsberger et~al.,}{Vogelsberger
  et~al.}{2014}]{vogelsberger2014introducing}
Vogelsberger M.,  et~al., 2014, MNRAS, 444, 1518

\bibitem[\protect\citeauthoryear{Wang, Li, Kauffmann  \& De~Lucia}{Wang
  et~al.}{2007}]{wang2007modelling}
Wang L.,  Li C.,  Kauffmann G.,   De~Lucia G.,  2007, MNRAS, 377, 1419

\bibitem[\protect\citeauthoryear{Weinmann, Kauffmann, Von Der~Linden  \&
  De~Lucia}{Weinmann et~al.}{2010}]{weinmann2010cluster}
Weinmann S.~M.,  Kauffmann G.,  Von Der~Linden A.,   De~Lucia G.,  2010, MNRAS,
  406, 2249

\bibitem[\protect\citeauthoryear{White \& Frenk}{White \&
  Frenk}{1991}]{white1991galaxy}
White S.~D.,  Frenk C.~S.,  1991, ApJ, 379, 52

\bibitem[\protect\citeauthoryear{Witten \& Frank}{Witten \&
  Frank}{2005}]{witten2005data}
Witten I.~H.,  Frank E.,  2005, Data Mining: Practical machine learning tools
  and techniques.
Morgan Kaufmann

\bibitem[\protect\citeauthoryear{Xu}{Xu}{1994}]{xu1994new}
Xu G.,  1994, arXiv preprint astro-ph/9409021

\bibitem[\protect\citeauthoryear{Xu, Ho, Trac, Schneider, Poczos  \&
  Ntampaka}{Xu et~al.}{2013}]{xu2013first}
Xu X.,  Ho S.,  Trac H.,  Schneider J.,  Poczos B.,   Ntampaka M.,  2013, ApJ,
  772, 147

\bibitem[\protect\citeauthoryear{Yoshida, Stoehr, Springel  \& White}{Yoshida
  et~al.}{2002}]{yoshida2002gas}
Yoshida N.,  Stoehr F.,  Springel V.,   White S.~D.,  2002, MNRAS, 335, 762

\makeatother
\end{thebibliography}
}

\appendix
\section{Feature Importance}
In the discussion of the halo properties chosen for our analysis, an evaluation of which attributes play a role in determining the galaxy properties was not performed. Here, we provide a feature imwportance plot that shows the relative importance of the halo properties (at $z=0$) in predicting galaxy properties.

\begin{figure}
\includegraphics[width=84mm]{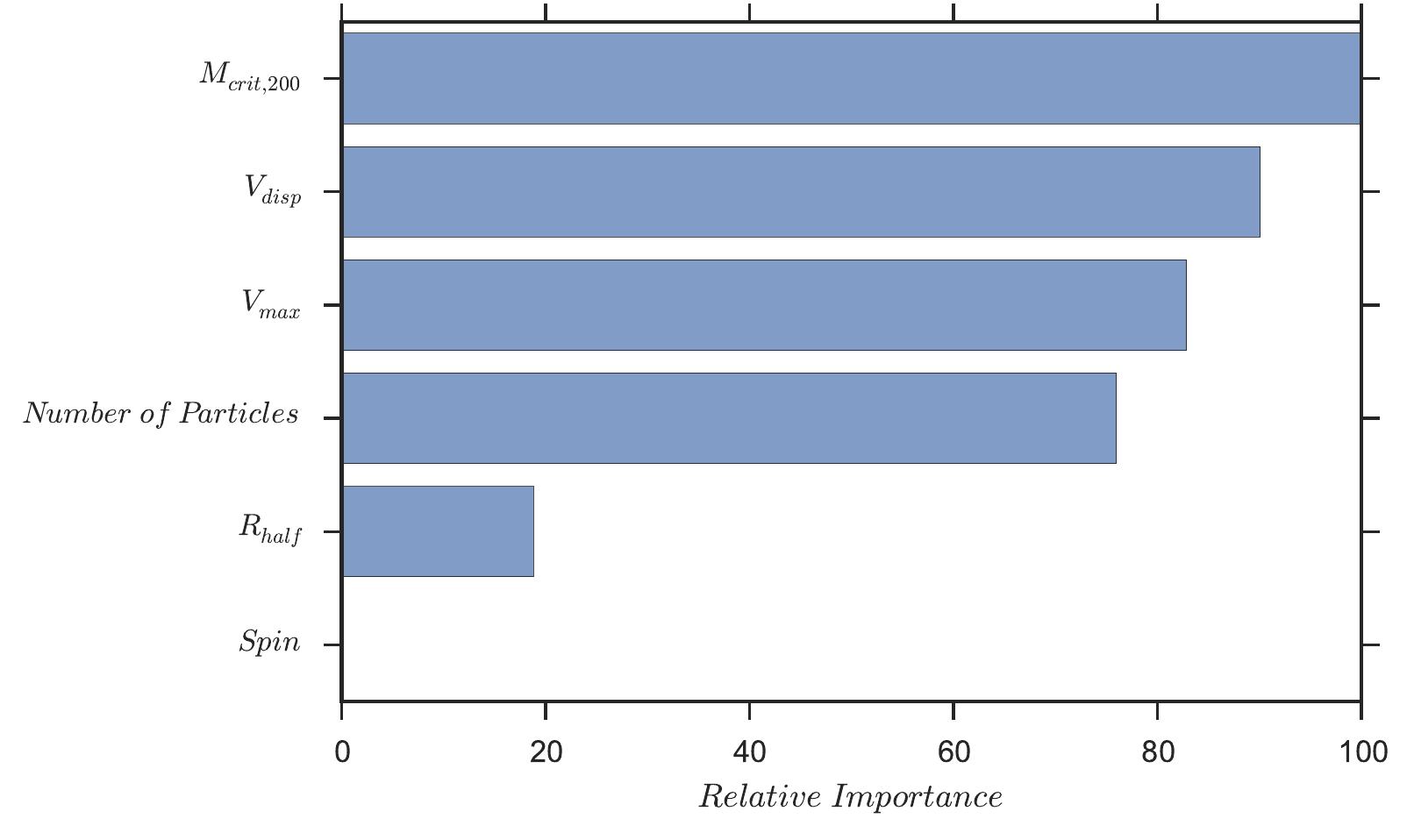}
\caption{The relative importance of different halo properties in predicting different properties of the galaxy.}
\label{feature}
\end{figure}

\par
For tree-based machine learning techniques, the depth of a feature (i.e. relative rank) used as a decision node can be used to evaluate how important that particular feature is in the learning process. The expected fraction of the samples a feature contributes to can be used as an estimate of the relative importance of the features. We then average this quantity over all trees in the ensemble to get a less biased estimate for the importance of a particular feature.
\par
As one would expect, the mass of the halo plays an integral role in determining the galaxy properties. Perhaps surprisingly, the spin of the dark matter halo plays a minimal role in the learning process. This analysis of feature importances will guide future work that uses machine learning to extract information from dark matter haloes about the galaxies residing in the halo.

\section{Using a different semi-analytical model}
An interesting question is whether machine learning techniques perform similarly well using a different SAM. The reason we used G11 for this work in place of, or along with, a Durham SAM \citep{bower2006breaking} was simply because more halo parameters were available in the merger trees that were constructed for DLB07 and G11. Using G11 offered the opportunity to explore a bigger parameter space.

Here, we explore the effect of using another SAM with fewer halo parameters. For \citet{bower2006breaking}, only the halo mass is provided through the merger tree. We repeat our analysis using just the halo mass over four snapshots and predict only the stellar mass and the black hole mass. The point of this analysis is to examine whether ML is able to model the same relationships when a different SAM is used with fewer inputs.

\begin{figure}
\includegraphics[width=84mm]{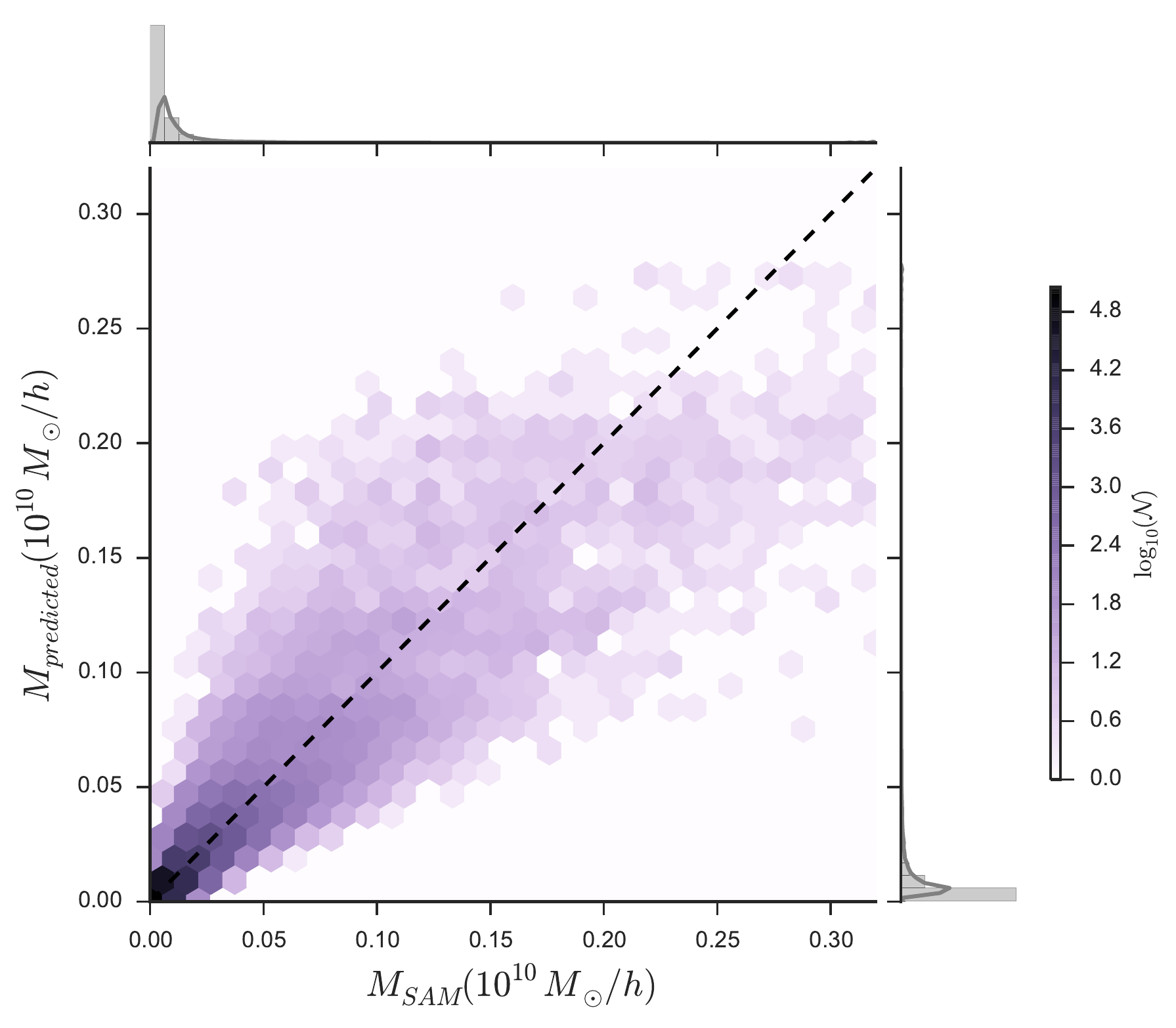}
\caption{A hexbin plot showing the black hole mass prediction for Bower et al. (2006) with a KDE on top.}
\label{bhbower}
\end{figure}

\begin{figure}
\includegraphics[width=84mm]{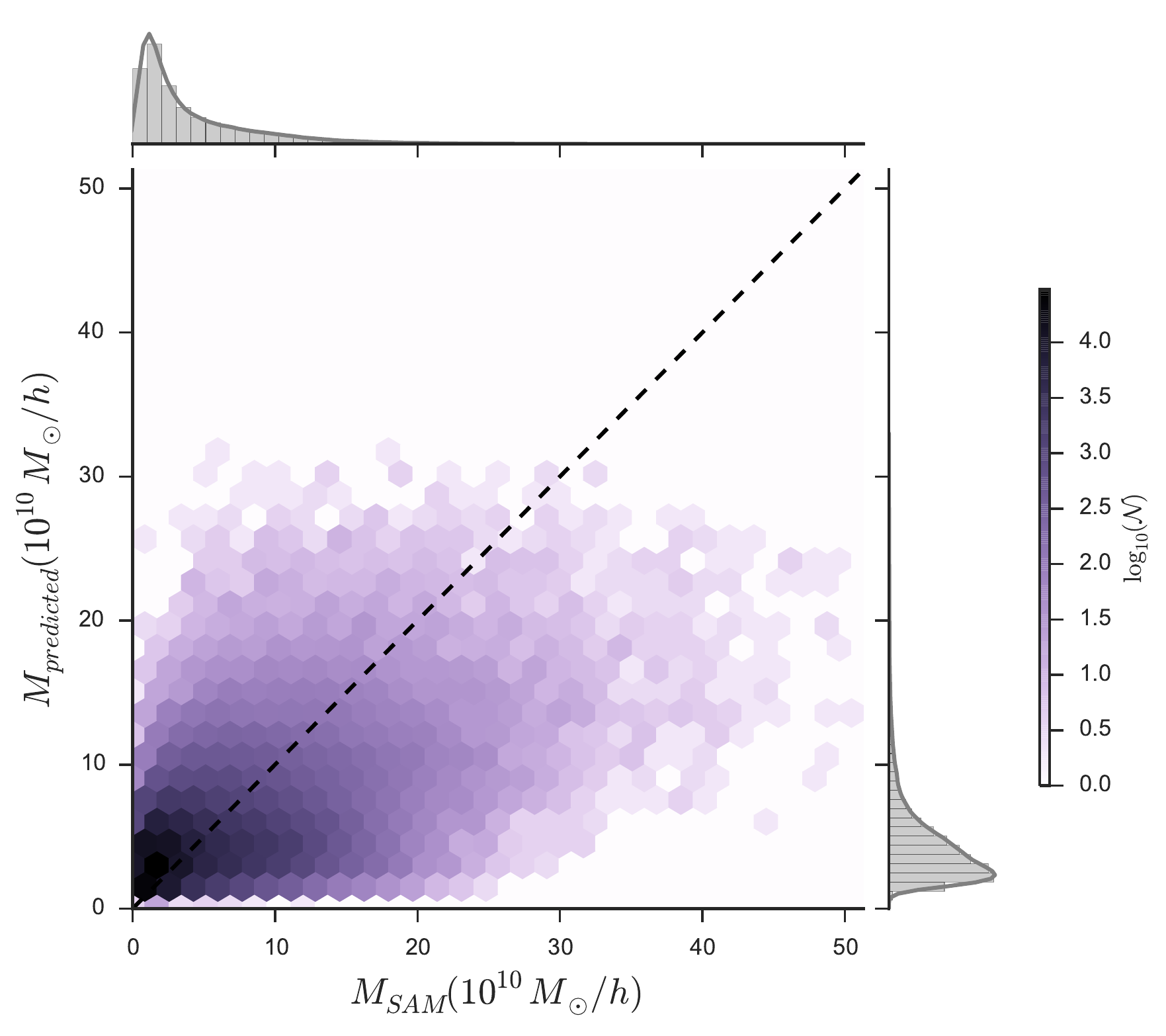}
\caption{A hexbin plot showing the stellar mass prediction for Bower et al. (2006) with a KDE on top.}
\label{stellarbower}
\end{figure}

\par
As we can see in the two figures attached, the predictions are noticeably more scattered (particularly the stellar mass) but the general trend is still recovered, even when only one feature is used in our prediction (out of necessity) in a SAM where some of the physics is treated differently. In Kamdar et al. (2015), we explore the feasibility of ML in making predictions from an NBHS, where the physics is vastly more complicated.
\\

\bsp
\label{lastpage}

\end{document}